\documentclass{aa}  

\usepackage{graphicx}
\usepackage{txfonts}
\usepackage{lipsum}
\usepackage{subcaption}         
\usepackage{lscape}             
\usepackage{placeins}           
                                
\usepackage{hyperref}
\usepackage{float}
\usepackage{multirow} 
\usepackage{arydshln}
\usepackage{xcolor}

\usepackage[normalem]{ulem}
\usepackage{orcidlink} 

\begin{document}

   \title{Insights into low-ionization structures in planetary nebulae: near-IR longslit spectroscopy}


   \author{M. B. Mari\orcidlink{0000-0002-2363-6568}\inst{1, }\inst{2}
        \and S. Akras\orcidlink{0000-0003-1351-7204}\inst{3}
        \and L. Konstantinou\orcidlink{0000-0002-1382-2367}\inst{3,}\inst{4,}\inst{5}
        \and D. R. Gonçalves\orcidlink{0000-0001-9388-7146}\inst{6}
        \and C. G. Díaz\orcidlink{0000-0001-5146-1358}\inst{1,2,7}
        }

   \institute{Universidad Nacional de Córdoba, Observatorio Astronómico de Córdoba, Laprida 854, X5000BGR Córdoba, Argentina\\
             \email{mbmari@unc.edu.ar}
             \and
            Consejo Nacional de Investigaciones Científicas y Técnicas (CONICET), Godoy Cruz 2290, CABA, CPC 1425FQB, Argentina
             \and
             Institute for Astronomy, Astrophysics, Space Applications and Remote Sensing, National Observatory of Athens, Penteli GR 15236, Greece
             \and
             European Southern Observatory (ESO), Alonso de Córdova 3107, Casilla 19, Santiago 19001, Chile
             \and
             Department of Physics, University of Patras, Patras, 26504 Rio, Greece
            \and
            Observat\'orio do Valongo, Universidade Federal do Rio de Janeiro, Ladeira Pedro Antonio 43, Rio de Janeiro 20080-090, Brazil
            \and
            Las Campanas Observatory, Carnegie Observatories, Casilla 601, La Serena, Chile
            }

   \date{Received xxx, 20XX}

 
  \abstract
   {Low-ionization structures (LISs) are small, dense clumps within planetary nebulae (PNe) that stand out in low-ionization emission lines, but the physical processes responsible for their excitation and the nature of H$_2$ remain poorly understood.}
   {We aim to characterize the excitation conditions and mass content of LISs in two PNe through near-IR longslit spectroscopic study of these structures, combining H$_2$, Br$\gamma$, and [Fe~{\sc ii}] diagnostics.} 
   {We obtained H- and K-band spectra with GNIRS at Gemini North for LISs in NGC~6818 and NGC~7354 --~objects with confirmed H$_2$ emission in their LISs. 
   We applied H$_2$ excitation diagrams, vibrational and rotational temperature diagnostics, classical near-IR diagnostic diagrams, and spatial profile analysis to constrain the physical conditions of the molecular and atomic gas phases.}
   {H$_2$, Br$\gamma$, and [Fe~{\sc ii}] emission are detected in all the structures studied. The H$_2$ level populations are consistent with predominantly thermal, near-LTE excitation, with excitation temperatures in the range $\sim$1860--2450~K. The [Fe~{\sc ii}]/Br$\gamma$ ratios are well below the values expected for shock- or X-ray-dominated regions, favoring photoionization as the dominant excitation mechanism of the atomic gas. A clear spatial stratification is observed in both objects, with Br$\gamma$ peaking closer to the central star and H$_2$~1--0\,S(1) and [Fe~{\sc ii}] co-spatial at larger distances. The warm molecular masses derived from H$_2$ emission lines are consistent with those estimated independently from H$_2$ column densities.
   }
   {The diagnostics presented here support a photoevaporation scenario in which LISs are dense clumps being progressively ionized from outside by the radiation of central stars, while their H$_2$ cores remain self-shielded. A slow-moving internal shock is likely responsible for releasing Fe into the gas phase. These results extend the molecular mass trend established previously to a larger sample. 
   }


   \maketitle
   \nolinenumbers

\section{Introduction}

Planetary nebulae (PNe) represent the ejected envelopes of low-to-intermediate mass stars during the final stages of their evolution before they become white dwarfs. Their morphologies are remarkably diverse --~exhibiting complex large-scale structures such as rims, shells and haloes, along with microstructures like filaments, jets and knots \citep{DeMarco2026}. Numerous dedicated works have studied the diversity and formation of PNe and their components; see \citet{2001ApJ...547..302G} and \citet{2002ARA&A..40..439B} for comprehensive reviews comparing models with observations, particularly in the optical, and addressing the main physical processes that govern the formation of both small- and large-scale nebular structures. Among this diversity, low-ionization structures (LISs, \citet{2001ApJ...547..302G}) are particularly intriguing: they are prominent in low-ionization emission lines such as [N~{\sc ii}], [S~{\sc ii}], [O~{\sc ii}] and [O~{\sc i}], and are kinematically and morphologically decoupled from the surrounding nebular shell. Understanding their origin and physical nature has been a long-standing challenge in the PNe community.

Despite nearly four decades of observational and theoretical effort \citep[][among others]{1987AJ.....94..671B,1996A&A...313..913C,Mellema1998,Odell2000,2001ApJ...547..302G,2009MNRAS.398.2166G,Odell2005,2008A&A...489.1141R,2016MNRAS.455..930A,2017MNRAS.465.1289A,2017PASA...34...36A,2023MNRAS.525.1998M,2023MNRAS.518.3908M,2025A&A...697A.227K,2025A&A...700A.155B}, the origin and formation mechanisms of LISs remain poorly understood. Proposed formation scenarios include collimated outflows driven by binary interactions, fossil condensations from the asymptotic giant branch (AGB) phase, and/or hydrodynamic instabilities in the expanding nebular shell \citep{1989MNRAS.241..625D,1998MNRAS.299..562S,1999ApJ...517..767G,2005ApJ...618..919G,2006ApJ...646L..61G,2022MNRAS.517.3822G,2025MNRAS.543.3867G,2001ApJ...556..823S,2008MNRAS.391.1063A,2020ApJ...889...13B,2022ApJ...933..168B}. In terms of excitation, the picture is similarly complex: contributions from both the hard radiation of the central stars and shock waves driven by fast outflows or local dynamical interactions, have been proposed. The relative importance of these mechanisms appears to vary among individual objects. This ambiguity is exemplified by the pairs of LISs in K~4-47 \citep{2000ApJ...535..823C,2004MNRAS.355...37G} and in NGC~7009 \citep{1998AJ....116..360B,2003ApJ...597..975G,2024A&A...689A..70A,2025A&A...700A.155B}, where spatially resolved spectroscopy has been crucial for advancing our understanding, although the dominant excitation mechanism remains debated.

Over the past few years, a series of discoveries has fundamentally changed our view of the composition of LISs and their physical nature. In particular, integral-field spectroscopy with MUSE at the VLT has revealed the presence of neutral carbon ([C~{\sc i}]~$\lambda$8727) in the LISs of NGC~7009 \citep{2024A&A...689A..70A} and in other PNe \citep{2022MNRAS.510.5444G,2025A&A...700A.155B}, tracing a transition layer between the ionized and molecular gas. The detection of emission from singly ionized iron and nickel ([Fe~{\sc ii}] and [Ni~{\sc ii}]) in LISs has added another piece to their perplexing picture \citep{2025A&A...700A.155B}. These species are indicative of shock-driven grain destruction, which releases iron and nickel from dust grains into the gas phase, pointing to an additional partially ionized zone at the surface of the LISs. These findings reveal a stratified structure within LISs, with low-ionization atomic lines giving way to neutral and then molecular tracers toward the interior, from the side exposed to the central star to the self-shielded core.
 
The molecular component of LISs has been confirmed through narrow-band H$_2$ imaging at 2.12~$\mu$m in a growing sample of PNe, both in pairs of LISs \citep{2015MNRAS.452.2445F,2018ApJ...859...92F,2017MNRAS.465.1289A,2020MNRAS.493.3800A,2025A&A...704A..59M} and in isolated structures \citep{2002ApJ...573L..55H,2009ApJ...700.1067M,2015ApJ...808..115M,2025MNRAS.541.1628B,2025A&A...704A..59M}.
A systematic analysis of this warm H$_2$ content by \citet{2025A&A...704A..59M} showed that the warm molecular mass of LISs ($\sim$5$\times$10$^{-7}$~M$_{\odot}$) is orders of magnitude smaller than their ionized mass ($\sim$7$\times$10$^{-4}$~M$_{\odot}$), suggesting that a largely unconstrained cold molecular reservoir may dominate their total mass. The next step is to move beyond imaging detections and exploit spectroscopy to characterize the excitation conditions of the molecular hydrogen, distinguishing between thermal and non-thermal processes, and probe the stratified physical structure of LISs layer by layer.

In this work, we present a near-IR longslit spectroscopic study of LISs, obtained with GNIRS at Gemini North for two PNe with confirmed H$_2$ emission in their LISs: NGC~6818 and NGC~7354. Section~\ref{obs} describes the observations and data reduction. Section~\ref{sect3} presents the detected emission lines. Section~\ref{DDs} analyzes the H$_2$ excitation diagnostics, and Section~\ref{sect_spatial} examines the spatial distribution of the emission lines. The results are discussed in Section~\ref{discussion} and summarized in Section~\ref{conclusions}.



\section{Observations}\label{obs}
To investigate the molecular hydrogen emission associated with LISs in greater detail, intermediate-dispersion near-infrared longslit spectra were obtained for two planetary nebulae, NGC 6818 and NGC 7354.

\begin{figure}[ht]
       \centering
       \includegraphics[width=0.8\columnwidth]{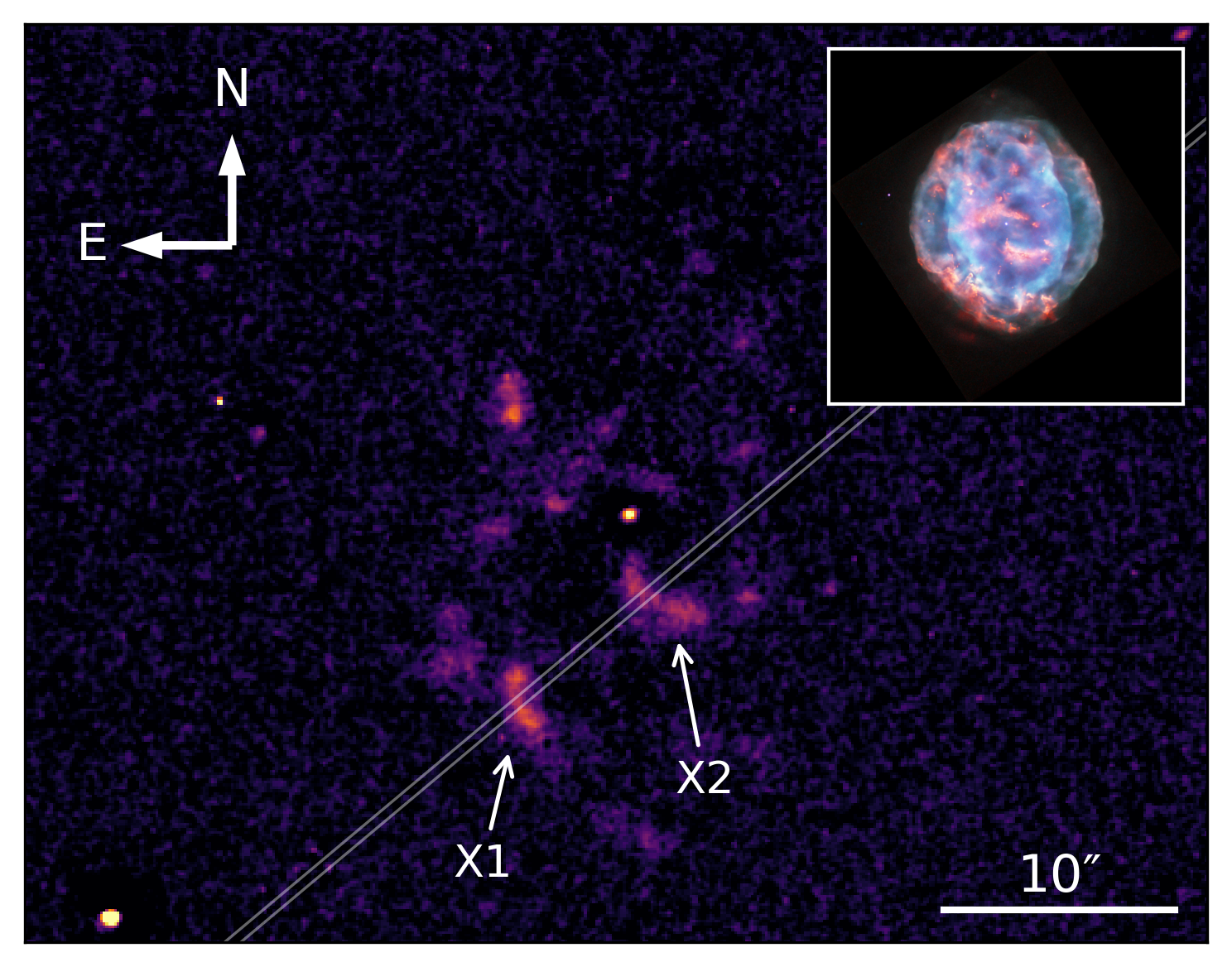}
       \includegraphics[angle=90,width=0.8\columnwidth]{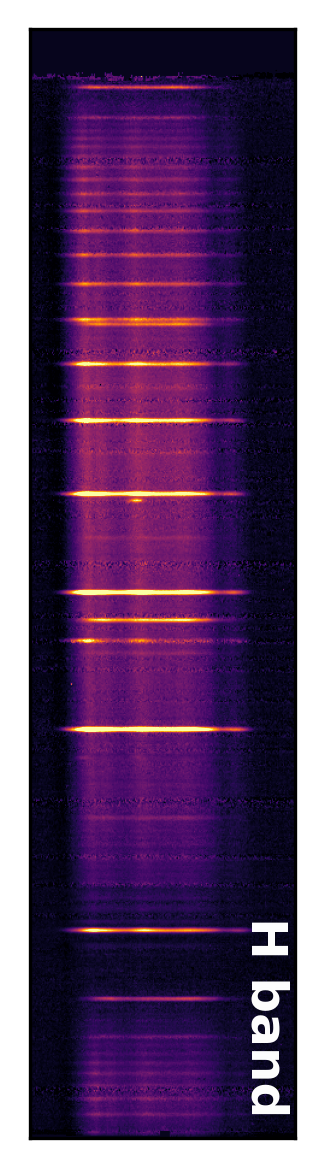}
       \includegraphics[angle=90,width=0.8\columnwidth]{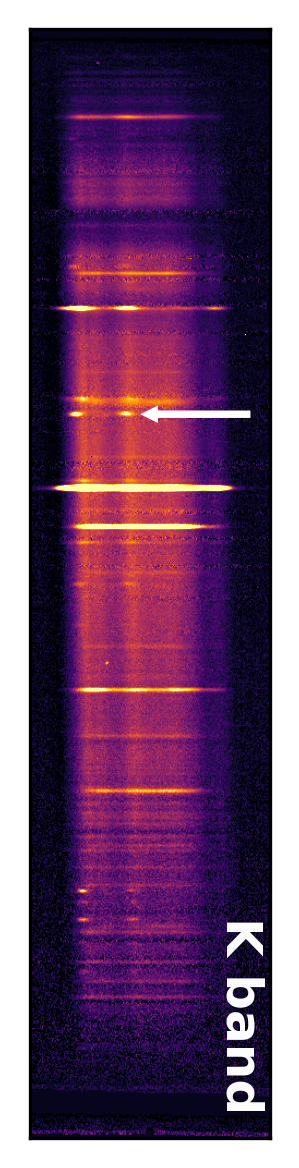}
       \caption{\textit{Top panel}: H$_2$~1-0~S(1) continuum-subtracted image of NGC~6818 \citep[][]{2025A&A...704A..59M}, with the slit overlaid at PA~=~130$^\circ$ and the regions studied in this work indicated. \textit{Middle panel}: Two-dimensional GNIRS spectrum along the slit in the H band. \textit{Bottom panel}: Same in the K band. In the bottom panel, the thick arrow marks the H$_2$~1-0~S(1) line at 2.122~$\mu$m, which reveals two spatially separated structures along the slit. An RGB HST image is also shown in the upper-right corner. Red colour corresponds to the [N~{\sc ii}]~$\lambda$6584 emission line.}
       \label{fig:1}%
\end{figure}

\begin{figure}[ht]
       \centering 
       \includegraphics[width=0.8\columnwidth,trim={0 0.25cm 0 0},clip]{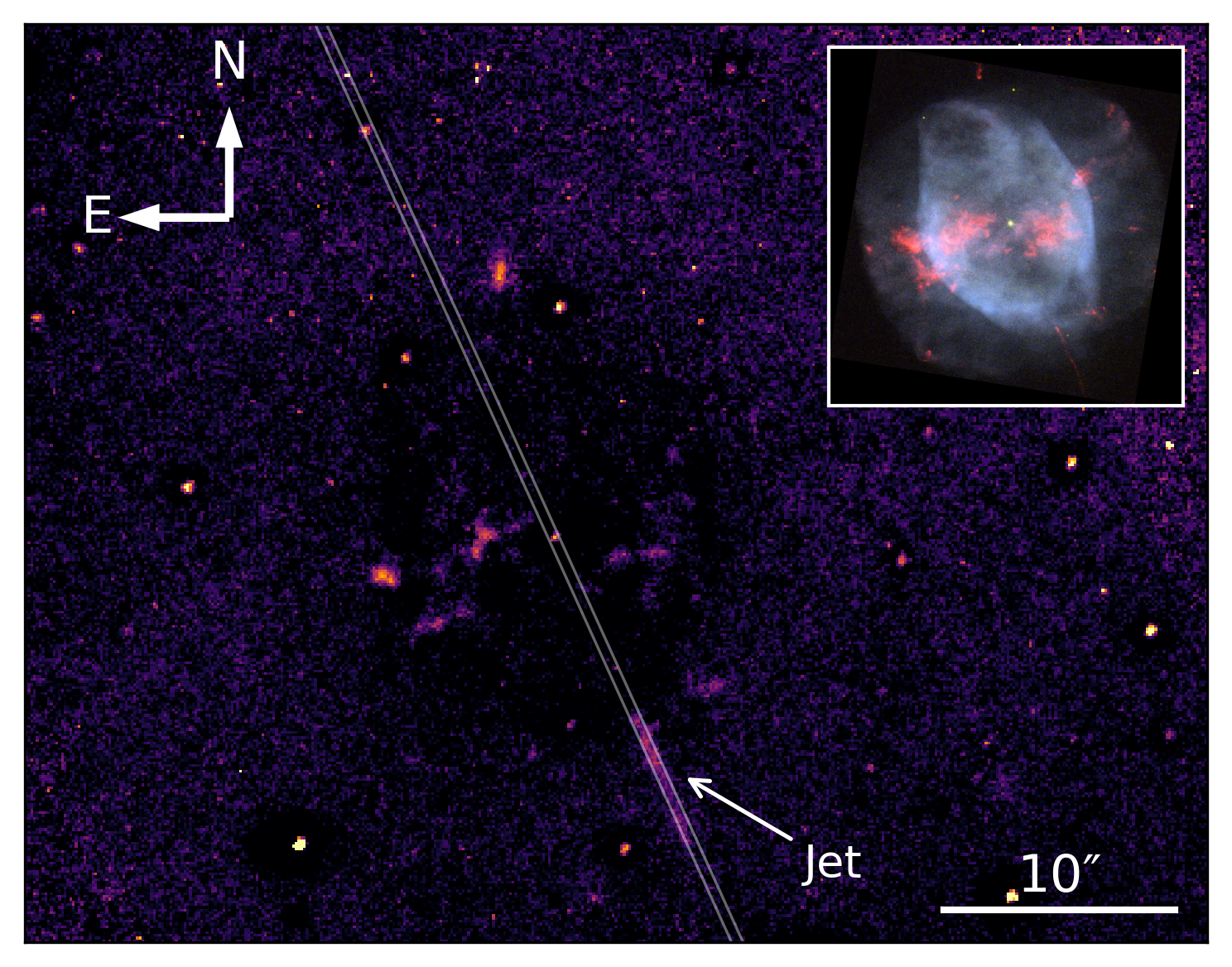}
       \includegraphics[angle=-90,width=0.8\columnwidth,trim={0.25cm 0 0 0},clip]{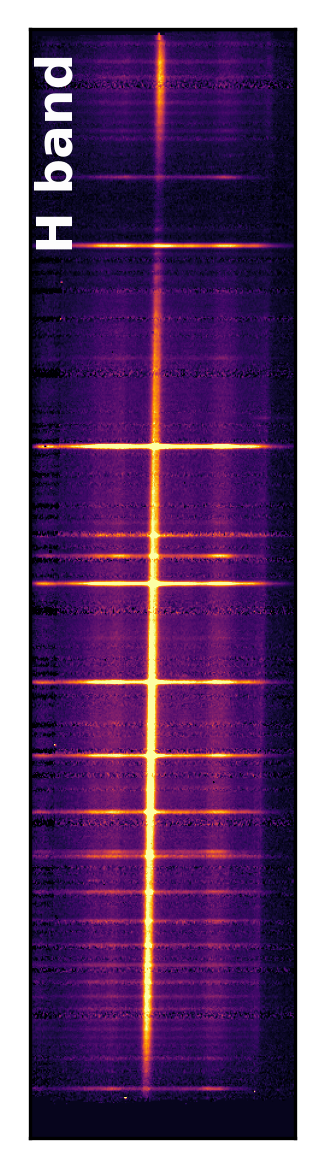}
       \includegraphics[angle=-90,width=0.8\columnwidth]{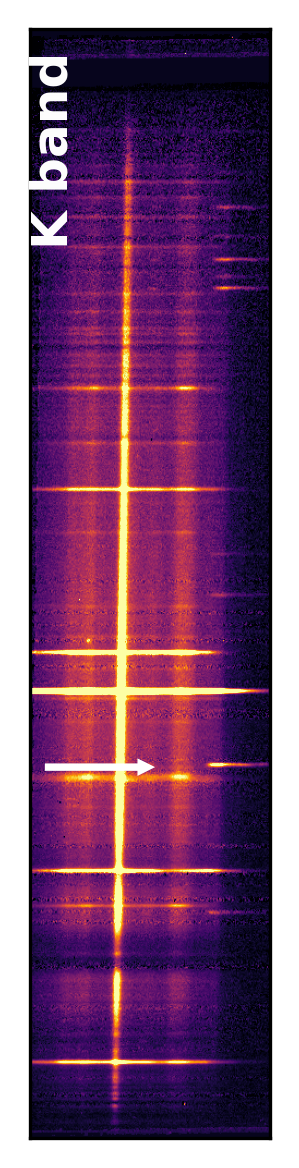}
       \caption{Same as Fig.~\ref{fig:1}, for NGC~7354. The slit position is overlaid at a position angle PA~=~24.4$^\circ$.}
       \label{fig:2}%
\end{figure}

The two PNe were selected from \citet{2025A&A...704A..59M}, where the presence of the H$_2$~1–0~S(1) emission line at 2.122~$\mu$m has been confirmed in several LISs from both objects. The upper panels in Figures~\ref{fig:1} and \ref{fig:2} display the H$_2$~1-0~S(1) continuum-subtracted images of the two PNe along with the position of the slits. The lower panels of the same figures show the two-dimensional spectra in the H- and K-bands, respectively. In both cases, the H${_2}~\lambda$2.122 emission line is indicated with an arrow.

The spectra in the H- and K-bands were obtained between May and June 2024 (Program ID: GN-2024A-Q-113, PI: M. B. Mari) using the Gemini Near-Infrared Spectrograph (GNIRS) mounted on the 8-m Gemini North telescope at Mauna Kea, Hawaii (see Table~\ref{tab:log}).

\begin{table}
\caption{Observing log. For each PN, the slit position angle (PA), observed band, individual exposure time, mean airmass, and number of science frames are listed. }
\label{tab:log}
\small
\centering
\setlength{\tabcolsep}{4pt}
\begin{tabular}{lccccc}
\hline
PN~Name   & PA     & Band  & Exp. time [s] &  Airmass  & \# Frames \\

\hline
\multirow{2}{*}{NGC~6818}   & \multirow{2}{*}{130$^{\circ}$}    &   H   & 180       & 1.209      & 16       \\
                            &                         &   K   & 180       & 1.632      & 12       \\
\hline
\multirow{2}{*}{NGC~7354}   & \multirow{2}{*}{24.4$^{\circ}$}   &   H   & 180       & 1.395      & 14       \\
                            &                         &   K   & 180       & 1.529      & 12       \\
\hline
\end{tabular}
%
\end{table}

Since the targets are spatially extended ($\sim$15–17 arcsec in radius), we used the short camera (0.15~arcsex~pix$^{-1}$) with a slit length of 99 arcsec, which is adequate for an ABBA dither pattern with $\pm$25~arcsec offsets along the slit. The 32~lines~mm$^{-1}$ grating and a 0.45~arcsec slit width were chosen in order to achieve an average resolving power of R$\approx$1100 in both bands. This configuration provides a reasonable sampling of the sky spectrum for an accurate subtraction. In agreement with the Gemini baseline calibrations, standard telluric stars (HIP 106737, HIP 103386, and FS 148), as well as spectroscopic flats and arcs (for each band), were also obtained.

\subsection{Data reduction}

First, as recommended by Gemini\footnote{\url{https://www.gemini.edu/instrumentation/gnirs/data-reduction}.}, we made use of the {\sc Python} routine \textit{CLEARIR.py} to remove artifacts such as vertical striping, quadrant-level offsets, and row-related noise.
The data reduction was made using the {\sc Gemini} package in {\sc Iraf}, following the standard procedure: flat field correction and wavelength calibration using lamp frames. Telluric correction was not necessary due to the absence of detectable stellar continuum in LISs. 
During the preparation of the raw data, the {\sc nsprepare} task generated variance and data quality extensions for each frame, enabling the propagation of uncertainties throughout the subsequent reduction procedures. This approach ensures that the final extracted spectra contain consistent estimates of the associated errors.

\subsection{Flux calibration}

Flux calibration was carried out by first retrieving the spectral types of the telluric standard stars 
(SIMBAD data), from which the corresponding effective temperatures were adopted to construct blackbody (BB) spectra. The H- and K-band transmission curves for 2MASS and GNIRS were obtained from the SVO Filter Profile Service\footnote{\url{https://svo2.cab.inta-csic.es/svo/theory/fps3/}}. The total BB flux transmitted through the 2MASS H and K bands were then estimated by integrating over the corresponding transmission curves, yielding model fluxes. 

The observed K-band magnitude of the telluric star was retrieved from the 2MASS catalog and converted to fluxes
using standard magnitude-to-flux relations. A scaling factor was then calculated by comparing the observed and model fluxes, and the BB spectrum was scaled accordingly. The scaled BB was interpolated over the GNIRS wavelength range, taking into account the GNIRS transmission curve.

A sensitivity function was constructed by dividing the scaled BB by the telluric star counts spectrum, which was applied to the science spectrum to obtain flux-calibrated data. 
The edge regions of the sensitivity function are less reliable because of the limits of the transmission curve. 

During this process, error propagation was also considered. The calibrated variance was computed by propagating both the object and sensitivity function uncertainties, and the resulting variance array was stored in the {\sc var} extension of the flux-calibrated {\sc fits} file to preserve the uncertainty information for subsequent analysis.

To compare the near-IR spectroscopic fluxes obtained with GNIRS with those derived from NIRI (Near-IR Imager, also mounted on the Gemini North telescope) imaging in  \citet{2025A&A...704A..59M}, Table~\ref{tab:comparison} presents a direct comparison between the two sets of measurements. All listed fluxes correspond to observed values, i.e. not dereddened.
The agreement is fairly good for the LISs in NGC~6818, despite the fact that the GNIRS slit samples only a fraction of the full transverse extent of these micro-structures. This suggests that uncertainties associated with continuum subtraction in narrow-band imaging may yield to a systematic underestimation of the H$_2$ fluxes derived from the narrow-band images. For the jet component in NGC 7354, where slit losses are expected to be smaller, the higher GNIRS flux likely reflects the combined effects of reduced continuum-subtraction uncertainties and the greater sensitivity of spectroscopy to faint H$_2$ emission. The complete set of all the measured line fluxes for both the H- and K-band spectra is provided in the Appendix~\ref{appendixA}.

\begin{table*}[ht]
\caption{Comparison between the LISs' H$_2$~1–0~S(1) fluxes measured from the NIRI images and those derived in this work from GNIRS. The nomenclature follows the labels introduced by \citet{2025A&A...704A..59M}, while the corresponding identifiers adopted in this work are given in parentheses.}
\label{tab:comparison}
\centering
\small
{
\begin{tabular}{llllcc}
\hline
Name                       & LISs      & RA   & Dec  & Flux 1-0~S(1) [NIRI]                           & Flux 1-0~S(1) [GNIRS]                         \\
                           &           &      &      & [$\times 10^{-15}$ erg s$^{-1}$cm$^{-2}$]    & [$\times 10^{-15}$ erg s$^{-1}$cm$^{-2}$]        \\
\hline                              
\multirow{2}{*}{NGC 6818}  &  2 (X1)   & 19:43:57.51  & $-$14:09:17.69  & 2.39         &  2.90$\pm$0.08       \\
                           &  3 (X2)   & 19:43:57.91  & $-$14:09:21.85  & 2.17         &  2.12$\pm$0.15      \\
                              
\hline
\multirow{1}{*}{NGC 7354}  & 3 (Jet) & 22:40:18.66 & $+$61:16:52.00  & 2.69        &  4.50$\pm$0.11      \\ 
\hline
\end{tabular}
}
\end{table*}


\section{Near-IR emission lines in LISs}\label{sect3}

\begin{figure*}[ht]
       \centering
       \includegraphics[width=0.75\textwidth]{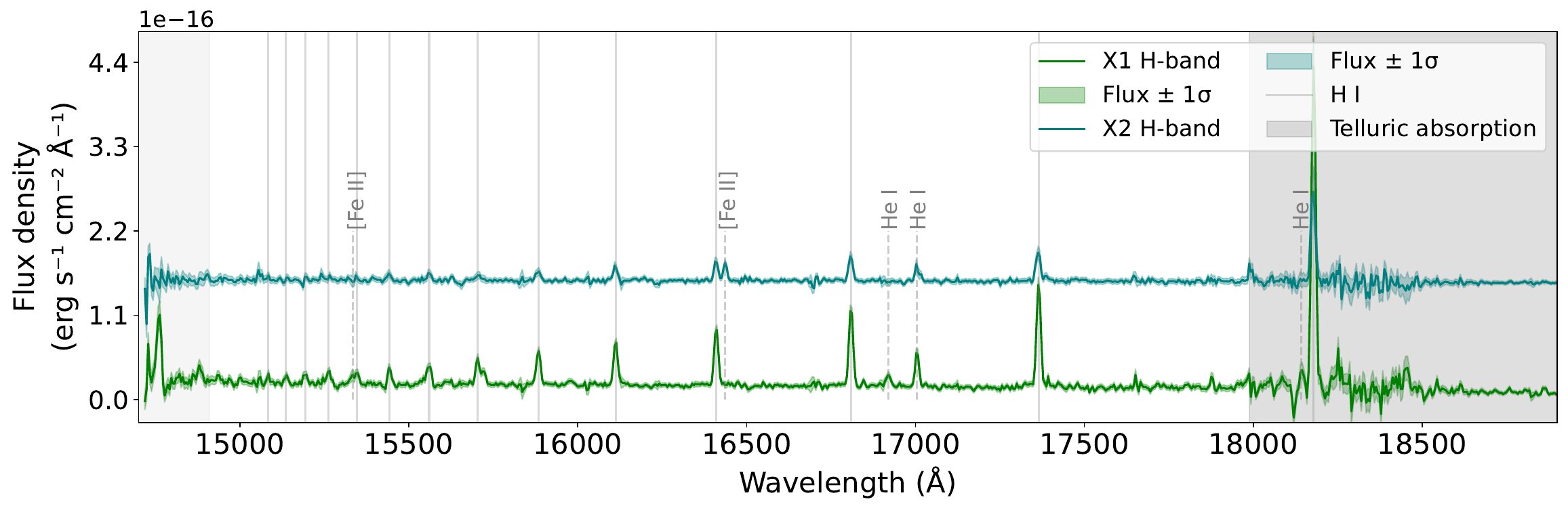}
       \includegraphics[width=0.75\textwidth]{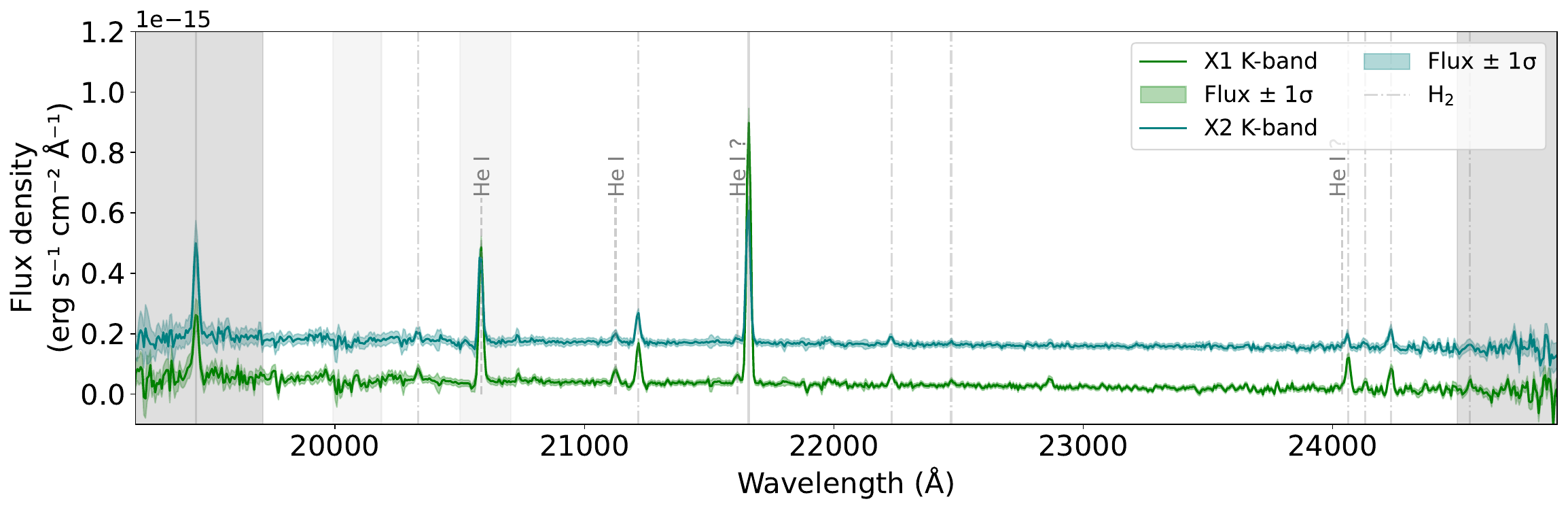}
       \caption{Extracted 1D spectra from the LISs in NGC 6818. The X2 spectrum is vertically shifted by $+1.5 \times 10^{-16}$ erg~s$^{-1}$cm$^{-2}$ $\text{\AA}^{-1}$ for clarity. The shaded bands around the spectra show the $\pm1\sigma$ uncertainty, which is narrow compared to the flux-density scale except near the spectral edges and in the telluric regions. Gray shaded regions mark telluric absorption regions, where flux measurements may carry increased systematic uncertainties. Darker gray shading marks regions of stronger absorption, while lighter shading denotes moderate telluric features \citep[][]{2015A&A...576A..77S}. In the H band, solid vertical lines indicate H~{\sc i} recombination transitions (Brackett series). In the K band, dash–dotted vertical lines mark H$_2$ emission lines from different rovibrational transitions.}
       \label{fig:1d_n6818}%
\end{figure*}

\begin{figure*}[ht]
       \centering
       \includegraphics[width=0.75\textwidth]{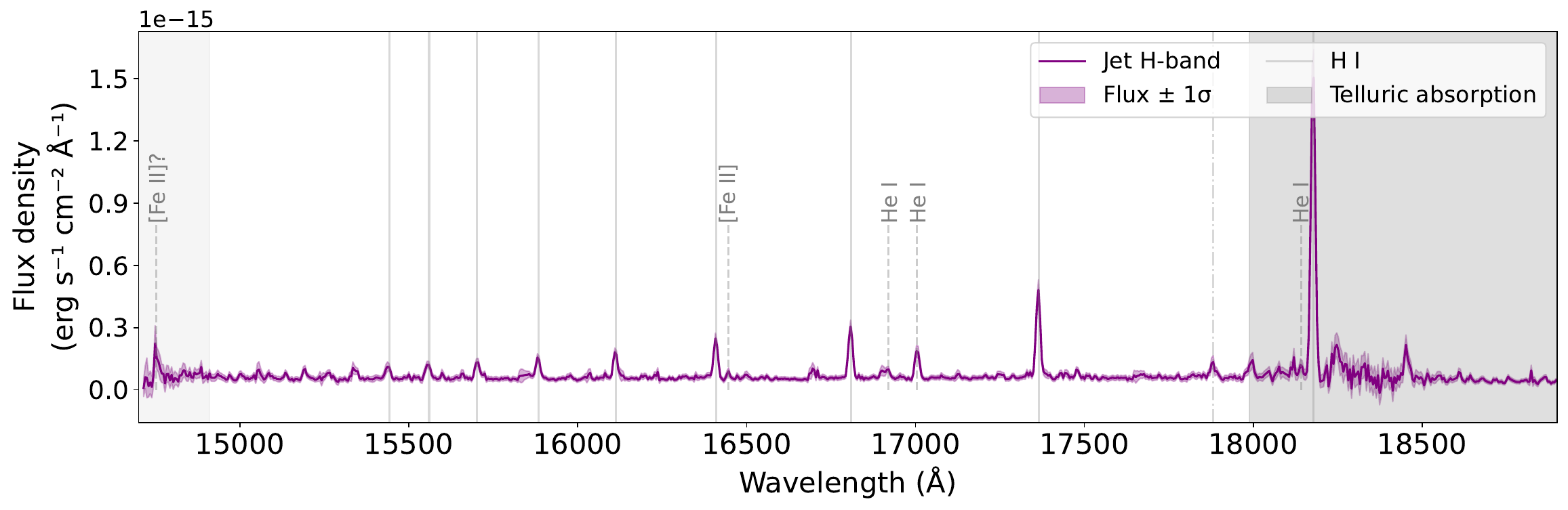}
       \includegraphics[width=0.75\textwidth]{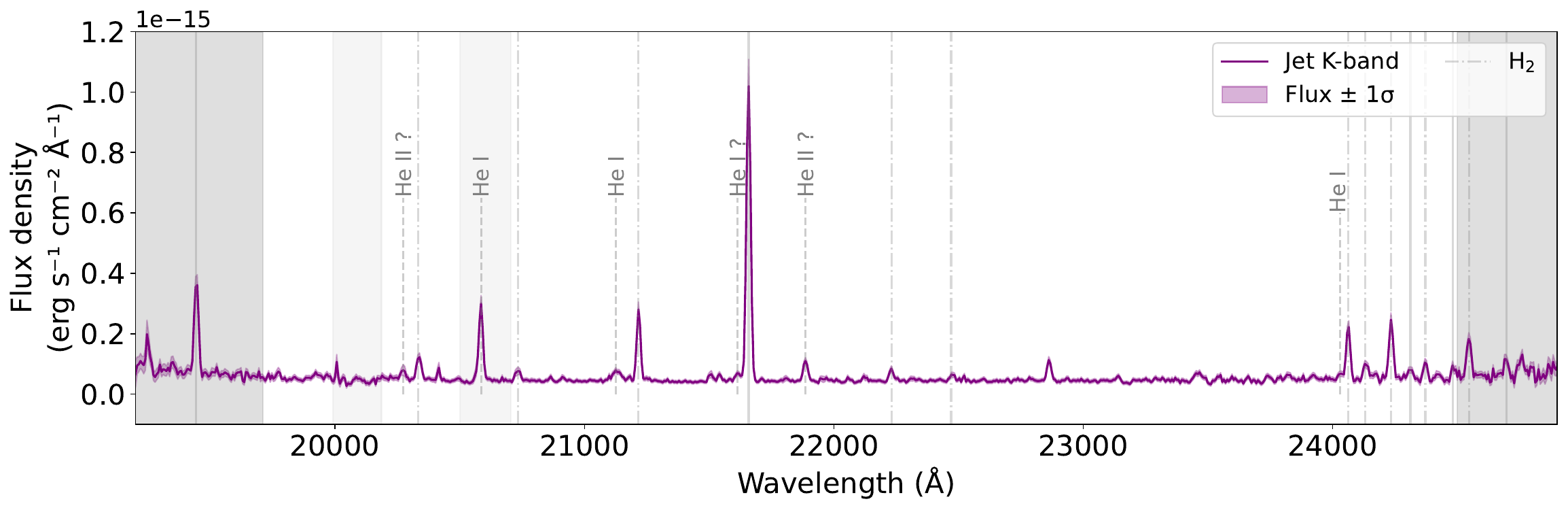}
       \caption{Extracted 1D spectra from the jet in NGC 7354. The shaded bands around the spectra show the $\pm1\sigma$ uncertainty, as in Fig.~\ref{fig:1d_n6818}. Gray shaded regions follow the same convention as in Fig.~\ref{fig:1d_n6818}. Solid vertical lines indicate H~{\sc i} recombination transitions (Brackett series), while dash–dotted vertical lines mark H$_2$ emission lines.}
       \label{fig:1d_n7354}%
\end{figure*}

Figures~\ref{fig:1d_n6818} and \ref{fig:1d_n7354} present the extracted H- and K-band spectra of the LISs X1 and X2 in NGC~6818, as well as the integrated spectrum of the jet in NGC~7354. These spectra are dominated by emission lines --~including several molecular hydrogen transitions, hydrogen recombination lines, and forbidden lines such as [Fe~{\sc ii}]~-- with no detectable \textbf{stellar} continuum, only a faint, nearly flat residual level ($\sim 5\times10^{-17}$~erg~s$^{-1}$~cm$^{-2}$~\AA$^{-1}$). For clarity, the spectrum of the X2 LIS in NGC~6818 is displayed with a vertical offset.

\begin{figure}[ht]
       \centering
       \includegraphics[width=0.59\columnwidth]{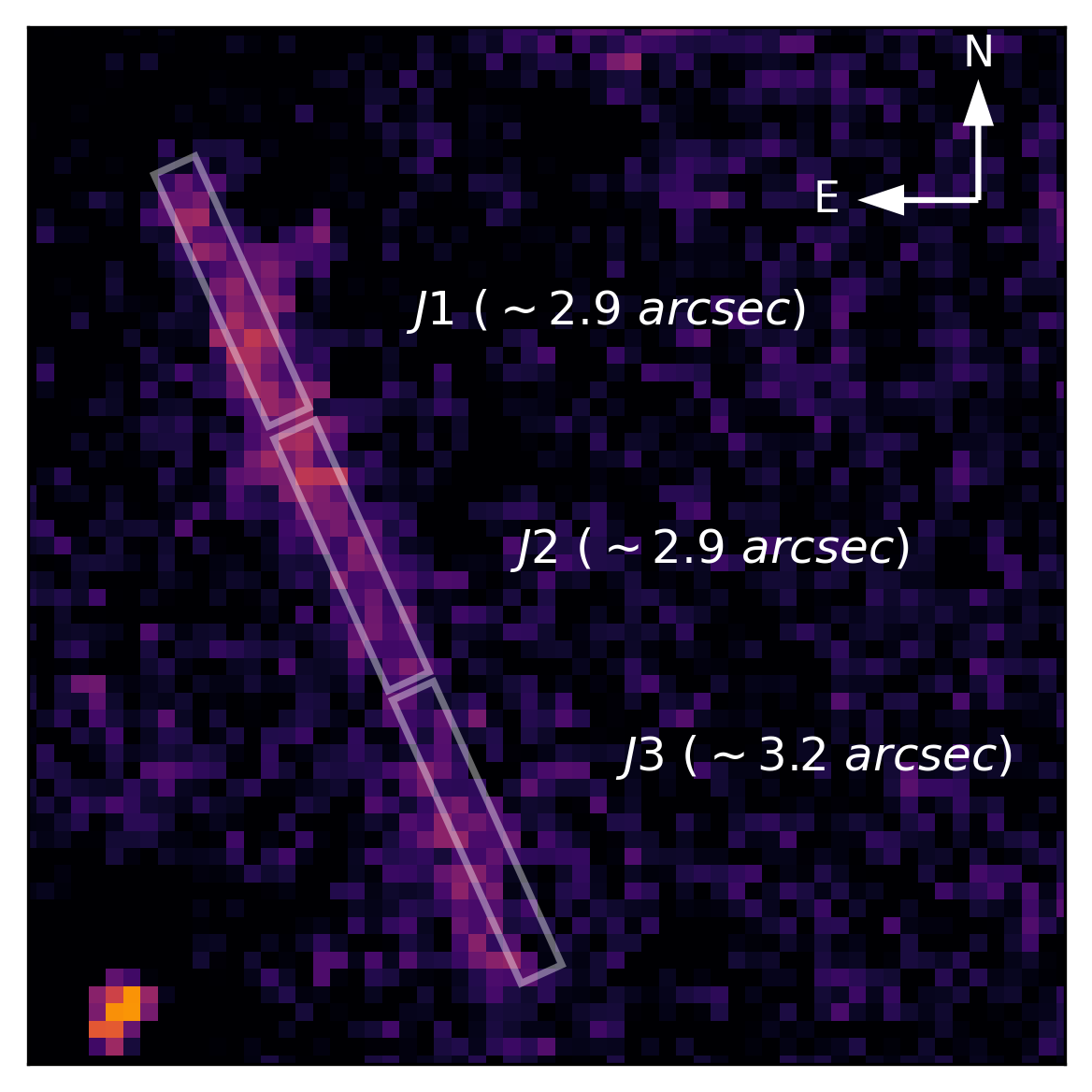}
       \includegraphics[width=0.2435\columnwidth]{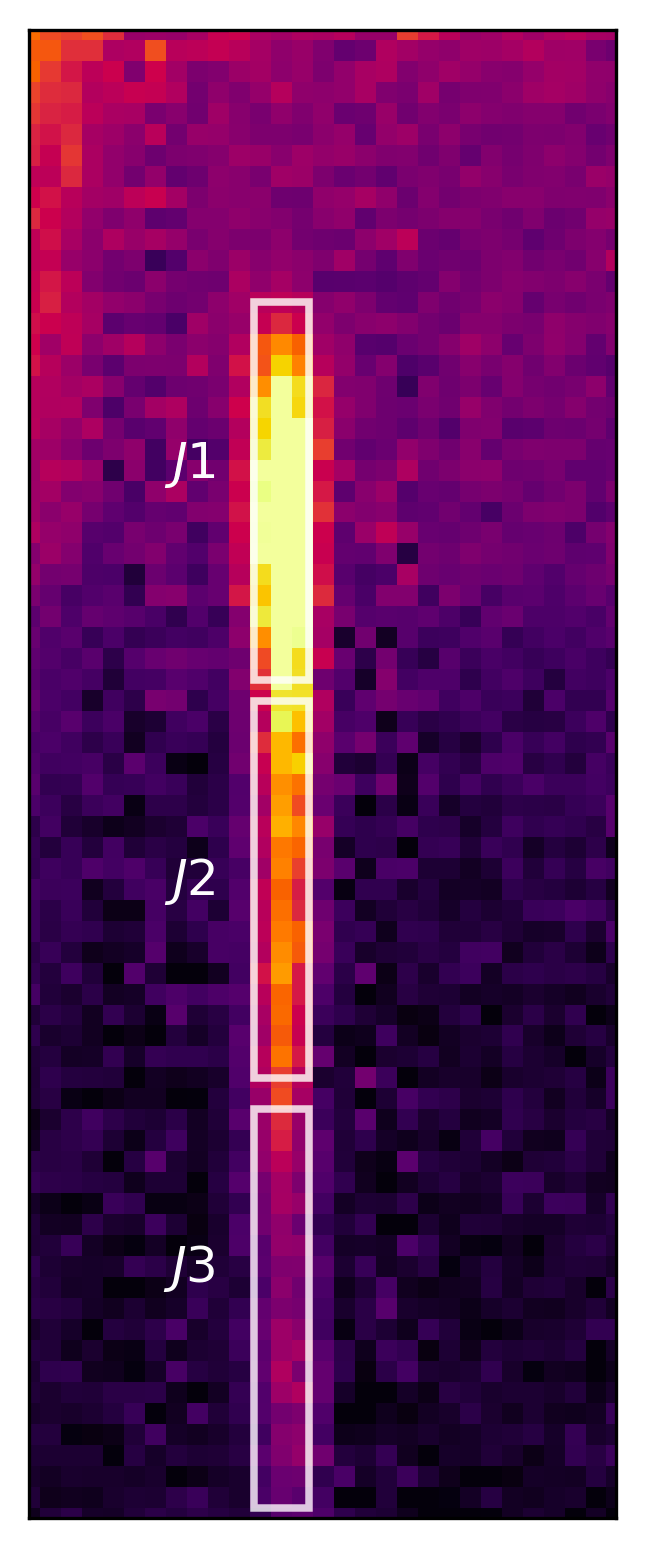}
       \caption{Zoomed-in views of the jet in NGC~7354 analysed in this work. \textit{Left panel}: Continuum-subtracted NIRI H$_2$ image, showing the full extent of the jet. \textit{Right panel}: two-dimensional GNIRS spectrum centred on the H$_2$ 1–0 S(1) line at 2.122~$\mu$m. The white rectangles mark the three jet segments: J1 (innermost), J2 (intermediate), and J3 (outermost), with projected lengths of $\sim$2.9,$\sim$2.9, and $\sim$3.2~arcsec, respectively.}
       \label{fig:jet_disection}%
\end{figure}

Taking advantage of the extended nature of the jet in NGC~7354 (total projected length $\sim$9~arcsec; see Fig.~\ref{fig:jet_disection}), we subdivided it into segments to investigate possible spatial variations along its length, for the first time in a planetary nebular jet. The jet was divided into three segments: J1 ($\sim$2.9~arcsec), the segment closest to the central star of the planetary nebula (CSPN); J2 ($\sim$2.9~arcsec), an intermediate segment; and J3 ($\sim$3.2~arcsec), the outermost segment. While the extracted spectra of these individual segments are not shown, their measured line fluxes (see Appendix~\ref{appendixA}) are used in the diagnostic diagrams and temperature estimates presented in the following sections. In addition, integrated fluxes for the entire jet are also measured and referred to as Jet.

The emission line fluxes were measured using the Line Measuring library \citep[{\sc LiMe},][version: 2.0.5]{2024A&A...688A..69F}, a {\sc Python}-based tool that assume Gaussian profiles. Taking into account the variance extension created during the reduction process, {\sc LiMe} estimates the uncertainties of the integrated fluxes through Monte Carlo algorithm with 1000 iterations\footnote{For further information about this tool and its error propagation, readers are directed to \url{https://lime-stable.readthedocs.io/en/latest/3_explanations/0_measurements.html   }}. The fluxes were measured after subtracting a local continuum, estimated for each line from adjacent line-free spectral bands. The spectra shown in Figs.~\ref{fig:1d_n6818} and \ref{fig:1d_n7354} are not continuum-subtracted, so the residual level does not affect the tabulated fluxes. 

To correct the emission lines for extinction, we adopted the literature values of the extinction coefficient (c$_{\beta}$): 0.37 for NGC~6818 \citep{2003A&A...400..161B} and 1.73 for the jet of NGC~7354 \citep{2010AJ....139.1426C}. Assuming a standard interstellar extinction law with R$_{V}=3.1$ \citep[][]{1989ApJ...345..245C}, and 
equations~(3) from the same work, the ratio $A({\rm H}\beta)/A_V\sim1.164$ was computed. Then, following the standard relation 
$A_V = c_{\beta}/(0.4\times A({\rm H}\beta)/A_V$) \citep[e.g.,][]{2006agna.book.....O}, the value of A$_V$ for each nebula was derived. Once A$_V$ was obtained, the equation~(2) from \citet{1989ApJ...345..245C}, valid in the range $0.909-3.333~\mu{\rm m}$ (that covers the H and K bands), was used to compute $A_{\lambda}/A_V$ for each line of interest. The dereddened fluxes were then calculated through $F_{\rm dered} = F_{\rm obs}\,10^{0.4 A_{\lambda}}$ (see Table~\ref{tab:dered}). 
While extinction correction substantially affects absolute fluxes (5–14\% for NGC~6818; 45–86\% for NGC~7354, strongest for [Fe~{\sc ii}]), its impact on the diagnostic ratios used here remains modest (4–6\%), except for [Fe~{\sc ii}]/Br$\gamma$ (5–20\%).

In addition to the larger extinction correction affecting [Fe~{\sc ii}] 1.644~$\mu$m, it is worth noting that this line is the only one in this analysis located in the H band, while all other near-IR lines used here (H$_2$ and Br$\gamma$) fall within the K band. Consequently, the [Fe~{\sc ii}]/Br$\gamma$ ratio is the only diagnostic that combines lines from different spectra, each with a distinct instrumental resolution: $\sim$13.6~\AA~ for the H band and $\sim$18.1~\AA~ for the K band, estimated from the average FWHM of arc-lamp lines in each spectrum. This additional source of variation, intrinsic to the cross-filter nature of the ratio, likely contributes to its comparatively larger sensitivity to the extinction correction.

\begin{table*}[ht]
\centering
\caption{Dereddened emission fluxes of H$_2$, Br$\gamma$, and [Fe~{\sc ii}] for different LISs. 
Fluxes were corrected using the \citet{1989ApJ...345..245C} extinction law. 
We adopted $c_\beta=0.37$ for the structures in NGC 6818 \citep{2003A&A...400..161B} 
and $c_\beta=1.73$ for the jet of NGC 7354 \citep{2010AJ....139.1426C}. 
Units: $[ \times 10^{-15} \mathrm{erg~s^{-1}~cm^{-2}} ]$.
}
\label{tab:dered}
\small
\begin{tabular}{lccccccc}
\hline
Neb & LIS & H$_2$ 2.0336 & H$_2$ 2.1222 & H$_2$ 2.2230 & H$_2$ 2.2474 & Br$\gamma$ 2.1657 & [Fe II] 1.6444 \\
\hline
\multirow[c]{3}{*}{NGC 6818} & & (A$_\lambda$=0.102) & (A$_\lambda$=0.096) & (A$_\lambda$=0.089) & (A$_\lambda$=0.087) & (A$_\lambda$=0.093) & (A$_\lambda$=0.144) \\
 & X1   & 0.70$\pm$0.06     & 3.17$\pm$0.09     & 0.84$\pm$0.13     & 0.36$\pm$0.07     & 19.84$\pm$0.23    & -- \\
 & X2   & 0.73$\pm$0.21     & 2.32$\pm$0.17     & 0.61$\pm$0.06     & 0.21$\pm$0.05     & 10.59$\pm$0.22    & 0.39$\pm$0.02 \\
 \hline
\multirow[c]{5}{*}{NGC 7354}  & & (A$_\lambda$=0.479) & (A$_\lambda$=0.447) & (A$_\lambda$=0.415) & (A$_\lambda$=0.408) & (A$_\lambda$=0.433) & (A$_\lambda$=0.674) \\
 & J1   & 1.20$\pm$0.11     & 3.41$\pm$0.07     & 0.68$\pm$0.07     & 0.38$\pm$0.05     & 21.82$\pm$0.17    & 0.42$\pm$0.09 \\
 & J2   & 1.03$\pm$0.10     & 2.49$\pm$0.08     & 0.70$\pm$0.06     & 0.36$\pm$0.05     & 6.96$\pm$0.08     & -- \\
 & J3   & 0.52$\pm$0.11     & 1.29$\pm$0.08     & 0.40$\pm$0.09     & 0.22$\pm$0.07     & 2.20$\pm$0.09     & --  \\
 & Jet  & 2.60$\pm$0.22     & 6.79$\pm$0.17     & 1.44$\pm$0.13     & 0.91$\pm$0.17     & 30.72$\pm$0.27    & 0.67$\pm$0.11 \\
 \hline
\end{tabular}
\end{table*}

\section{\texorpdfstring{H$_2$ excitation mechanism diagnostics}{H2 excitation mechanism diagnostics}}\label{DDs}

There are two main mechanisms responsible for the excitation of molecular hydrogen: (i) collisional excitation (thermal processes) and (ii) radiative excitation followed by cascade decay from the excited electronic states (non-thermal processes). These mechanisms produce distinct spectral signatures, which are often observed simultaneously within a single object.

In the thermal case, H$_2$ molecules are excited through inelastic collisions with atoms or molecules in warm gas (with kinetic temperatures typically in the range T$_{\rm kin}\sim$500-2000~K), heated by shocks \citep[e.g.,][]{1989MNRAS.236..929B}, X-rays \citep[e.g.,][]{1996ApJ...466..561M} or UV photons in high-density regions \citep[ $>$10$^{5}$~cm$^{-3}$,][]{1989ApJ...338..197S}. Under these conditions, to first approximation, the H$_2$-emitting gas is in local thermal equilibrium (LTE), meaning that a single excitation temperature (T$_{\rm ex}$) fully characterizes its thermodynamic state (where the vibrational and rotational temperatures are approximately equal). Consequently, the resulting near-IR spectrum can be described by a single T$_{\rm ex}$, typically below $\sim$4000~K, since H$_2$ is rapidly dissociated by thermal collisions above this temperature.

In non-thermal excitation, near-IR ro-vibrational lines arise from radiative cascades through the vibrational and rotational levels of the ground electronic state. The excitation is produced by the absorption of a UV photon in the Lyman–Werner bands (912–1108\AA) or by collisions with fast electrons generated by X-ray ionization \citep[][]{1987ApJ...322..412B}.  This non-thermal excitation mechanisms typically occurs in cold gas \citep[T<100--200~K][]{1987ApJ...322..412B} and leads to level populations that deviate significantly from LTE.

\subsection{\texorpdfstring{H$_2$ lines as diagnostic tools}{H2 lines as diagnostic tools}}\label{sect:h2diag}

The relative intensities --~the extinction-corrected fluxes~-- of H$_2$ emission lines can be used to diagnose the dominant excitation mechanism. In particular, the ratio H$_2$~2–1S(1)/1–0S(1) is an excellent discriminator between non-thermal and thermal processes, while H$_2$~1–0S(2)/1–0S(0) helps to distinguish between thermal UV excitation and X-ray or shock heating \citep[][]{1994ApJ...427..777M,2013MNRAS.428.2389M}. These ratios, calculated from the integrated emission over the full spatial extent of each LIS, are also sensitive to the vibrational (T$_{\rm vib}$) and rotational (T$_{\rm rot}$) temperatures. 
Assuming the spontaneous radiative decay probabilities A$_{\rm \nu\nu' JJ'}$ from \citet{1977ApJS...35..281T}, and following the expressions derived by \citet{2013MNRAS.428.2389M}, we adopt the following relations for the vibrational and rotational temperatures:

$$T_{\text{vib}} \simeq \frac{5600}{\ln\left(1.355 \times \frac{I_{1-0S(1)}}{I_{2-1S(1)}}\right)} ;\hspace{0.5cm} T_{\text{rot}} \simeq \frac{-1113}{\ln\left(0.323 \times \frac{I_{1-0S(2)}}{I_{1-0S(0)}}\right)}$$

\begin{figure}[ht]
       \centering
       \includegraphics[width=0.97\columnwidth]{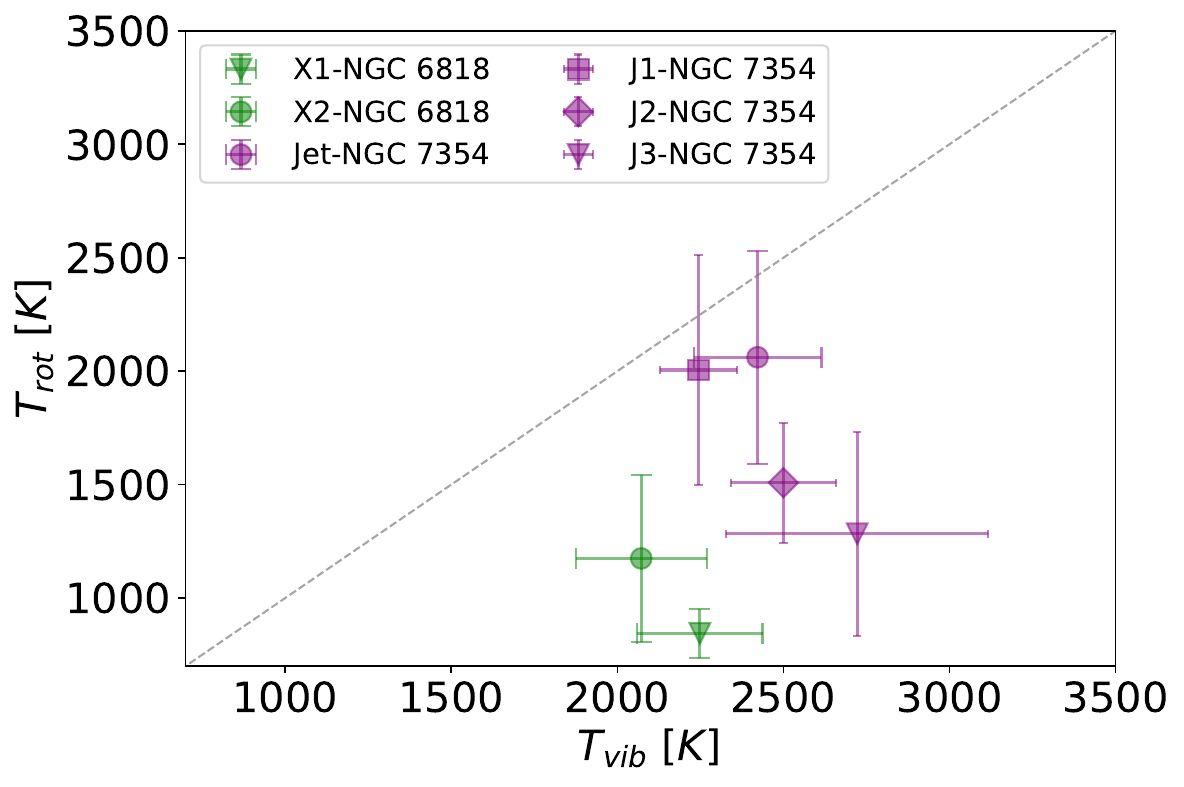}
       \caption{Vibrational (T$_{\rm vib}$) and rotational (T$_{\rm rot}$) excitation temperatures derived from H$_2$ line ratios for the LISs in NGC~6818 and the jet segments (J1–J3) and integrated jet in NGC~7354. The dashed line indicates equality between both temperatures.}
       \label{fig:Tvib_Trot}%
\end{figure}

Figure~\ref{fig:Tvib_Trot} shows that, within the uncertainties, the vibrational and rotational temperatures are relatively consistent for all analyzed structures. However, the structures in both PNe systematically lie below the identity line, indicating that \(T_{\rm rot}\) is lower than \(T_{\rm vib}\). According to \citet{2013MNRAS.428.2389M}, this behavior is consistent with a dominant thermal excitation mechanism with a secondary contribution from non-thermal processes, such as UV fluorescence or X-ray heating. This motivates the use of additional diagnostics to further investigate the possible contribution of non-thermal processes.

\begin{figure*}[ht]
     \centering
            \includegraphics[width =0.33\textwidth]{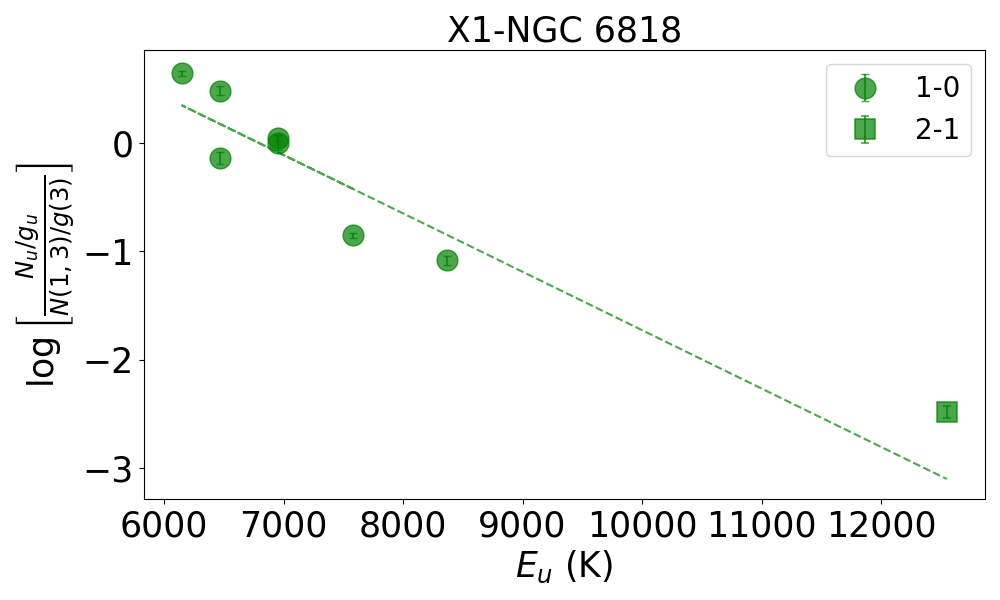}
            \includegraphics[width =0.33\textwidth]{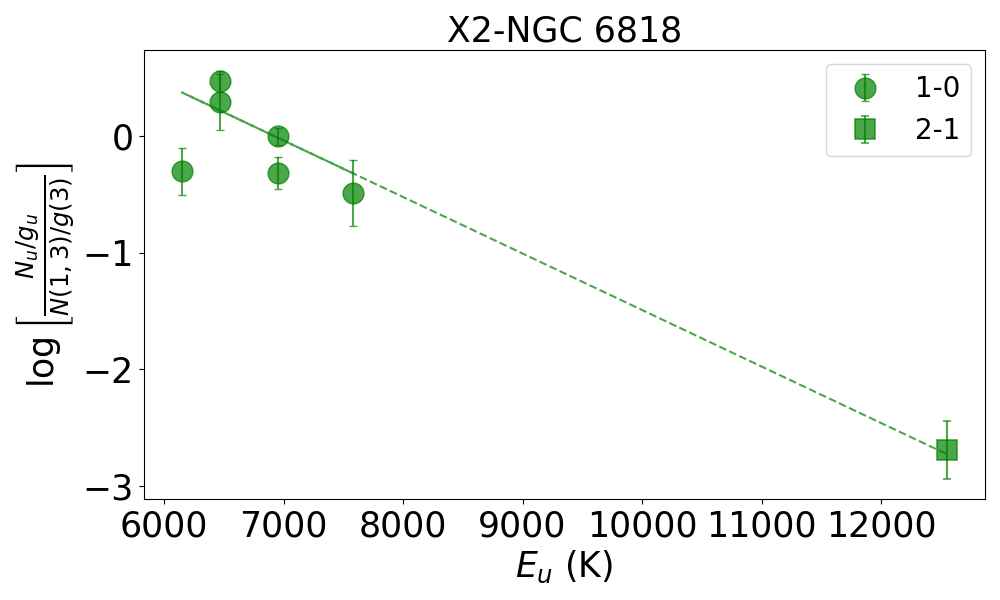}            
            \includegraphics[width=0.33\textwidth]{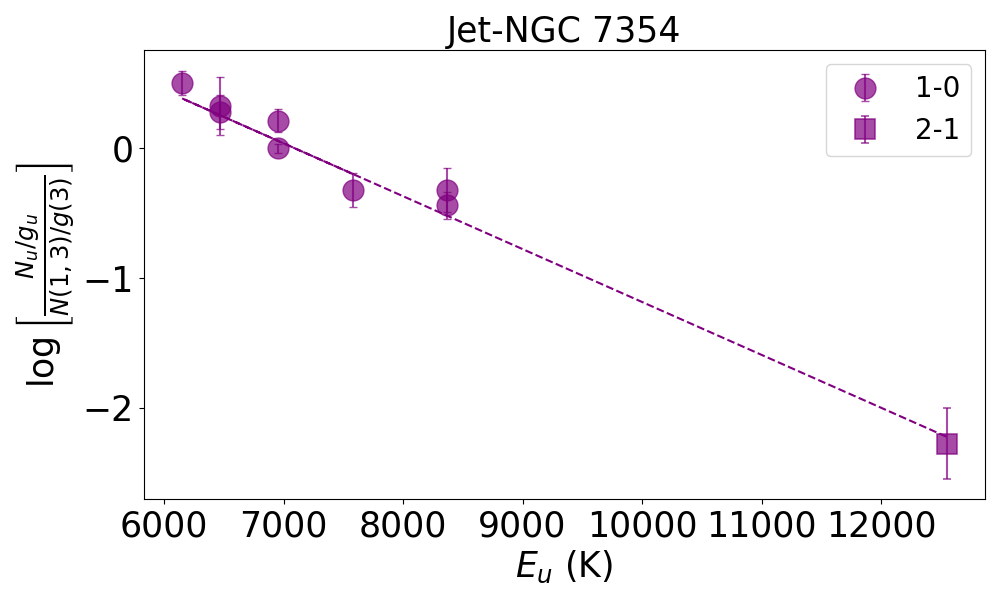}
            \includegraphics[width=0.33\textwidth]{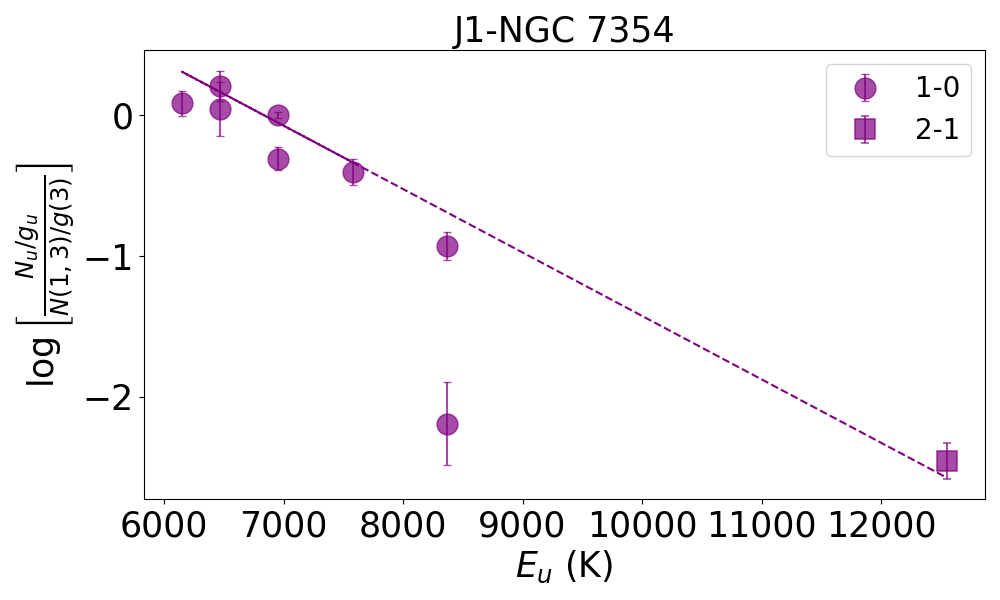}
            \includegraphics[width=0.33\textwidth]{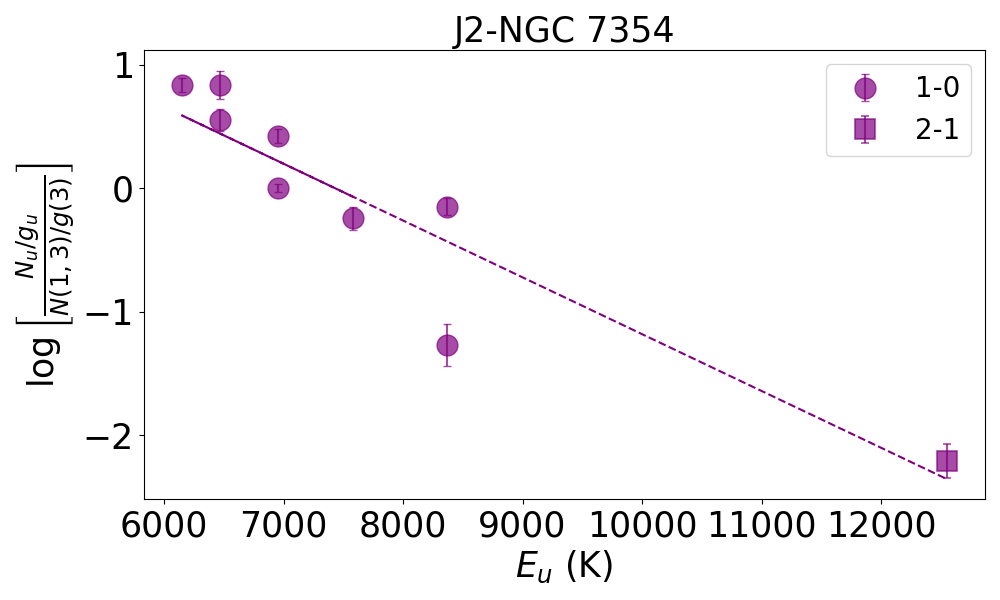}
            \includegraphics[width=0.33\textwidth]{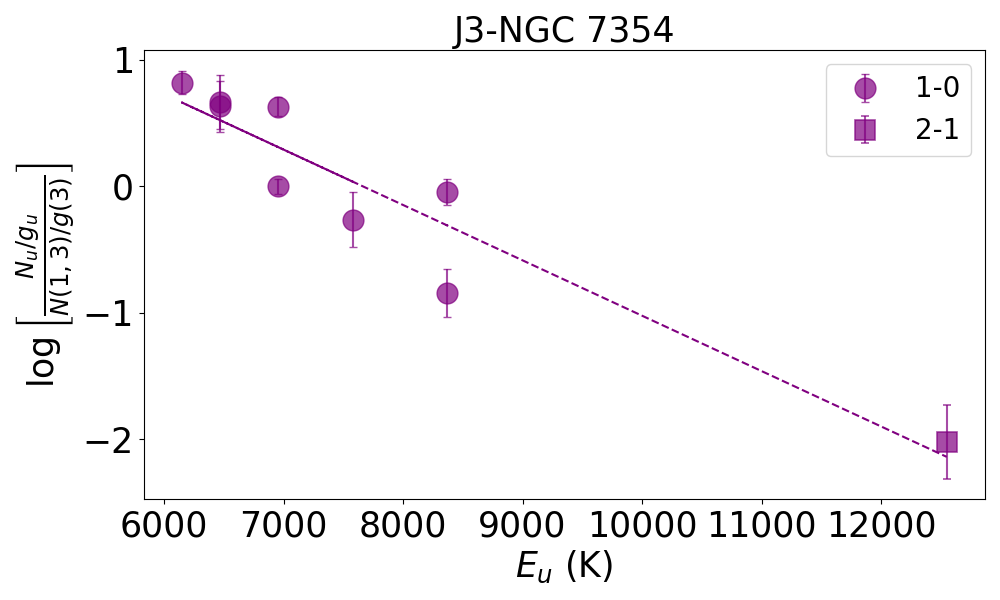}
    \caption{Excitation diagrams of the regions X1 and X2 in NGC~6818, the Jet in NGC~7354 and the three jet segments J1, J2 and J3.} 
    \label{excitation_plots}
\end{figure*}

Another approach to assessing the excitation conditions of H$_2$ is to examine the population of its energy levels using excitation diagrams.
These diagrams show the column density of the upper level normalized by its statistical weight, N$_u$(J)/g$_u$(J), as a function of the upper level energy E$_u$(K). 
To estimate N$_u$(J) we employed the measures H$_2$ emission line fluxes and the transition probabilities from \cite{Wolniewicz_1998}, while E$_u$(K) was retrieved from \citet{Roueff2019}.
In LTE conditions, the level populations follow a Boltzmann distribution and are well described by a linear fit, a behavior commonly associated with collisional excitation induced by shocks or by collisions within a dense gas. 
Deviations from linearity, on the other hand, are indicative of UV excitation \citep[e.g.,][]{2015MNRAS.453.1888M}. However, since we detect only H$_2$ emission lines arising from the vibrational transitions v~=~1 $\rightarrow$ 0 and v~=~2 $\rightarrow$ 1, our excitation diagrams do not include levels with upper vibrational states (v$>$3), where the non-thermal signature of UV excitation would emerge most clearly. The H$_2$ lines were fitted using a weighted least-squares method, with weights determined from the uncertainties in the emission line fluxes. During the fitting process the H$_2$ 2-1 S(3), H$_2$ 1-0 S(6) and H$_2$ 1-0 S(5) transitions were removed. The first one was excluded because it was significantly affected by residual sky-subtraction noise, while the final two lines were removed since they lie near the edge of the H band, where the measurements are more uncertain.
Figure~\ref{excitation_plots} show that the H$_2$ lines lie along a straight line in all the analized structures, consistent with LTE conditions. 
The inverse of the slope yields the excitation temperature T$_{\rm ex}$, while the intercept provides the total column density of molecular hydrogen N(H$_2$), once the partition function is taken into account. 
The derived values are listed in Table~\ref{table_TexNH2}.
With the exception of X1 in NGC~6818, all the structures show T$_{\rm ex}>2000$~K, and the column densities are comparable to those found for the cometary knots in the Helix nebula \citep{2007MNRAS.382.1447M} and the LISs of NGC~7009 (Konstantinou et al., submitted).
The derived excitation temperatures T$_{\rm ex}$ are in reasonable agreement with the vibrational temperatures T$_{\rm vib}$ estimated independently from H$_2$ line ratios (Table~\ref{table_TexNH2}), with differences generally below $\sim$20\%. 
In contrast, the rotational temperatures T$_{\rm rot}$ are systematically lower than both T$_{\rm ex}$ and T$_{\rm vib}$ across all structures, consistent with the trend already noted in  Fig.~\ref{fig:Tvib_Trot}. This overall agreement between T$_{\rm ex}$ and T$_{\rm vib}$ further supports the conclusion that the H$_2$ emission arises under near-LTE conditions, while the lower T$_{\rm rot}$ values suggest a secondary contribution from non-thermal processes, as discussed above.

\begin{table*}[ht]
\centering
\small
\caption{Excitation temperatures (T$_{\rm ex}$) and H$_2$ column densities (N(H$_2$)) derived from excitation diagrams (Fig.~\ref{excitation_plots}) together with vibrational (T$_{\rm vib}$) and rotational (T$_{\rm rot}$) temperatures estimated from H$_2$ line ratios.}
\label{table_TexNH2}

\begin{tabular}{lccccc}  
        \hline
        PN & LIS & T$_{\rm ex}$ [K]& N(H$_2$) [$\times$ 10$^{16}$ cm$^{-2}$]   & T$_{\rm vib}$ [K] &  T$_{\rm rot}$ [K] \\
        \hline
    \multirow[c]{2}{*}{NGC 6818}    & X1 & 1\,860 $\pm$ 50 & 27.40 $\pm$ 1.60    & 2\,250$\pm$190 & 840$\pm$110   \\
        & X2 &  2\,060 $\pm$ 180  & 13.20 $\pm$ 3.40 & 2\,070$\pm$200 & 1\,170$\pm$370  \\
        \hline
    \multirow[c]{4}{*}{NGC 7354}    & J1 &  2\,220 $\pm$ 100 & 12.6 $\pm$ 1.60   & 2\,240$\pm$120 & 2\,010$\pm$510  \\
        & J2 & 2\,180 $\pm$ 100  & 9.20 $\pm$ 1.30    & 2\,500$\pm$160 & 1\,510$\pm$270  \\
        & J3 & 2\,290 $\pm$ 200 & 4.00 $\pm$ 0.30    & 2\,720$\pm$390 & 1\,280$\pm$450  \\
        & Jet & 2\,450 $\pm$ 210 & 4.75 $\pm$ 0.60   & 2\,420$\pm$190 & 2\,060$\pm$470  \\

        \hline
        
    \end{tabular}
\end{table*}

\begin{figure}[ht]
       \centering
       \includegraphics[width=0.98\columnwidth]{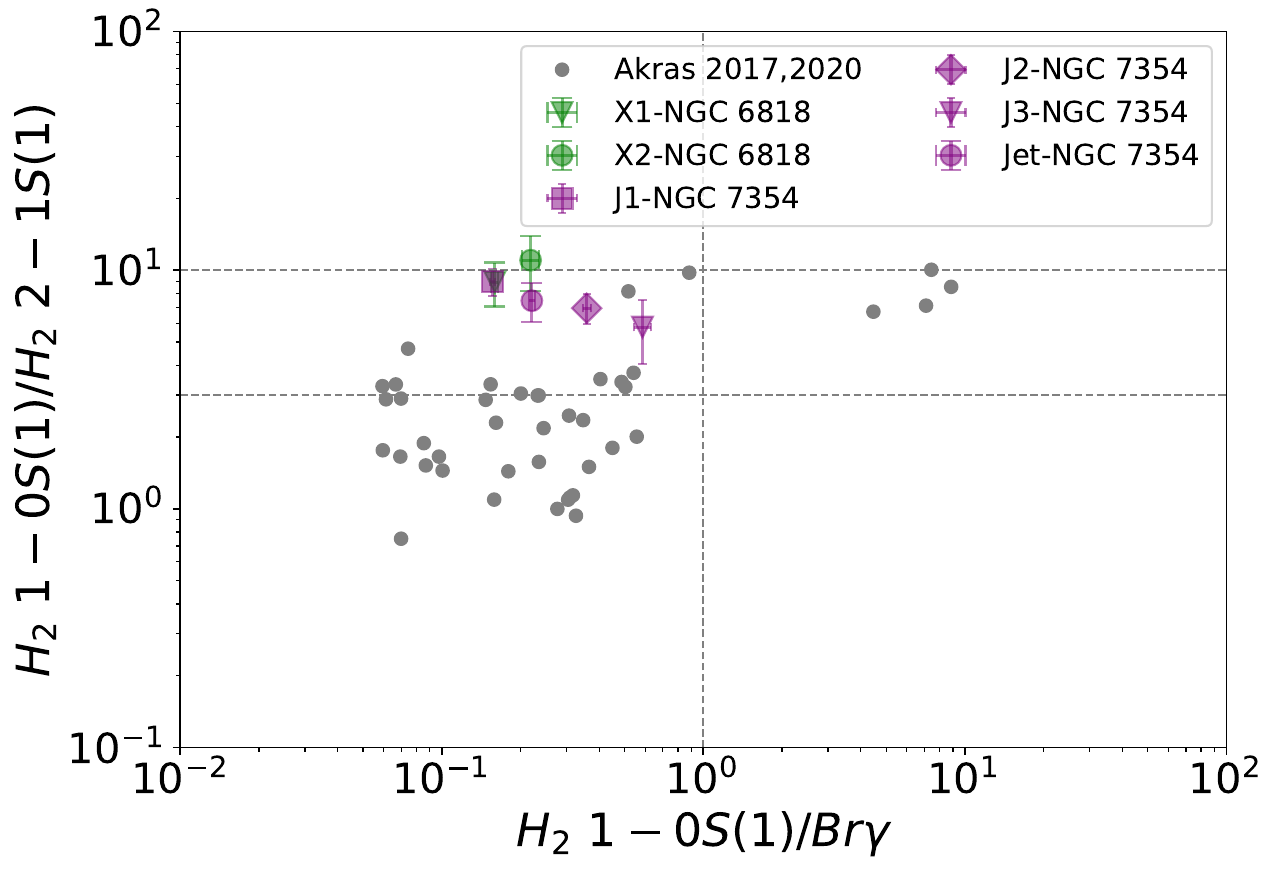}
       \caption{R(H$_2$)=H$_2$1–0 S(1)/H$_2$2–1 S(1) and R(Br$\gamma$)=H$_2$1–0 S(1)/Br$\gamma$ DD. The symbols represent the LISs and jet components analyzed in this work, while literature values from \citet{2017MNRAS.465.1289A,2020MNRAS.493.3800A} are shown for comparison in gray points. The horizontal lines delimit the regions associated with UV excitation ($<$3) and shock excitation ($>$10), while the vertical one separates Br$\gamma$-bright sources (to the left) from H$_2$-bright sources (to the right) as adopted by \citet[][]{2015MNRAS.453.1888M}.}
       \label{fig:H2_H2_akras}%
\end{figure}

Another quantity that is useful as a diagnostic of the H$_2$ excitation mechanism is the line ratio R(H$_2$)=H$_2$~1--0S(1)/2--1S(1) \citep[][]{1980AJ.....85..886B}. When combined with R(Br$\gamma$)=H$_2$~1--0S(1)/Br$\gamma$, this ratio provides a widely used diagnostic to distinguish between UV-dominated and shock-dominated excitation regimes. Using this R(Br$\gamma$) versus R(H$_2$) diagram, \citet{2015MNRAS.453.1888M} showed that sources with strong H$_2$ emission tend to occupy the shock-dominated region, whereas UV-excited sources generally display Br$\gamma$ emission brighter than H$_2$~1--0S(1). 
More recently, \citet{2020MNRAS.493.3800A} reported a subtle correlation in this diagnostic diagram (DD), with both ratios increasing from the lower right to the upper left. Nevertheless, they concluded that both excitation mechanisms can reasonably reproduce the observed range of R(H$_2$) values. Figure~\ref{fig:H2_H2_akras} shows the location of the LISs analyzed in this work, together with values compiled from \citet{2017MNRAS.465.1289A,2020MNRAS.493.3800A}.

The R(Br$\gamma$) versus R(H$_2$) diagram provides an empirical view of H$_2$ excitation regimes but does not uniquely distinguish thermal from non-thermal excitation.
It is also important to note that at high densities ($n>10^{5}~{\rm cm}^{-3}$)  collisional de-excitation can thermalize the low-energy states of H$_2$ molecules initially excited by UV photons \citep[][]{1989ApJ...338..197S,Burton1990}. As a result, the derived line ratios may be close to those expected from pure shock-excitation \citep[][]{1995ApJ...455..133H}. Therefore, the apparent consistency between T$_{\rm vib}$ and T$_{\rm rot}$ should not be interpreted as unambiguous evidence of shock excitation, since H$_2$ excited by UV photons can also exhibit quasi-thermal level populations when the gas density is sufficiently high.
Estimates of the total hydrogen density in these structures, obtained from the empirical gas-to-extinction relation of \citet{1978ApJ...224..132B} (see Appendix~\ref{appendixB}), yield $n_{\rm total}\sim1.7$--$7.6\times10^{4}$~cm$^{-3}$; adopting instead the slit width as a proxy for 
the (unresolved) transverse depth of the jet, $n_{\rm total}$ reaches $\sim5\times10^{5}$~cm$^{-3}$, within the density regime required for the thermalization scenario discussed above.

\begin{figure*}[ht]
       \centering
       \includegraphics[width=0.98\columnwidth]{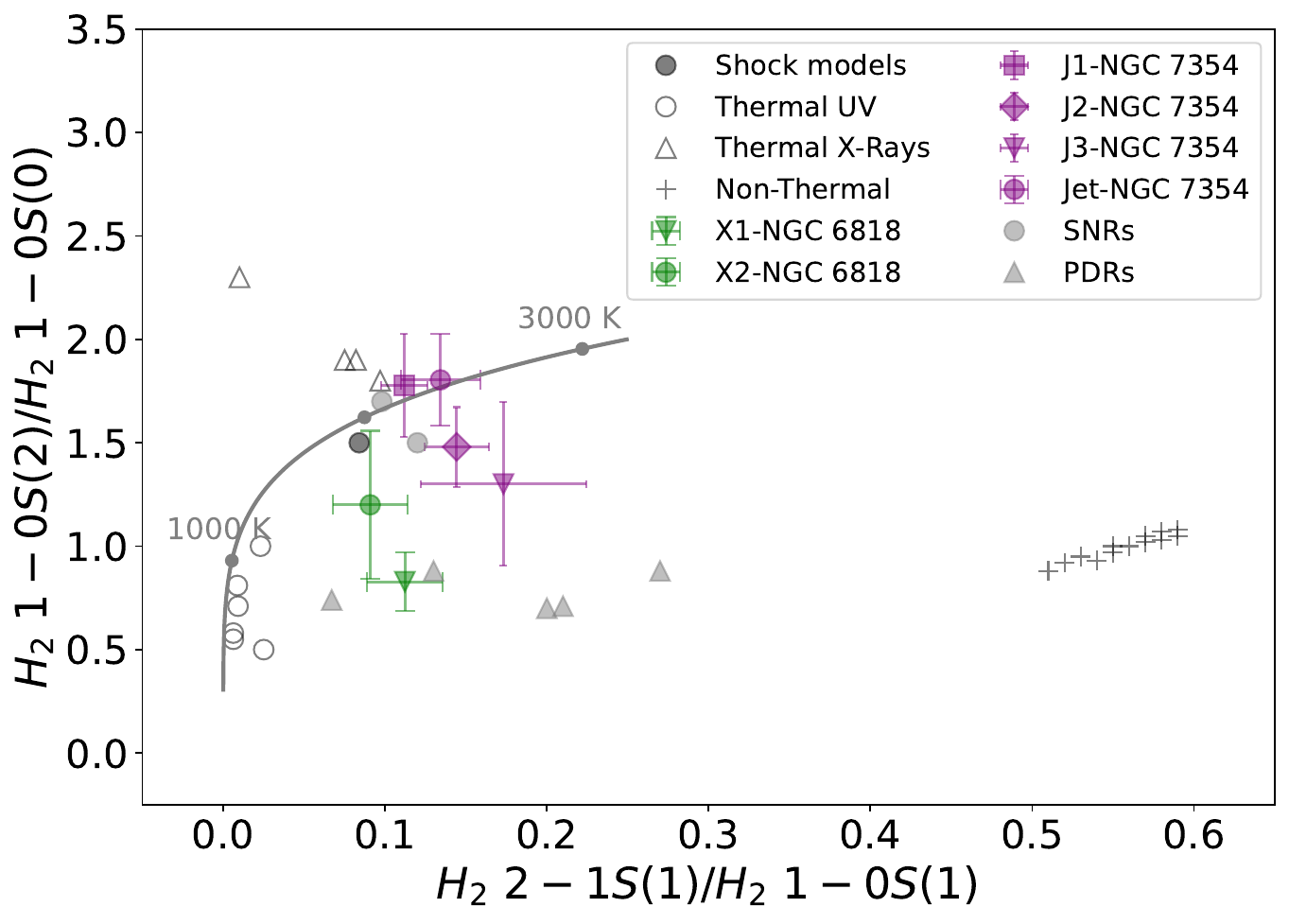}
       \includegraphics[width=0.98\columnwidth]{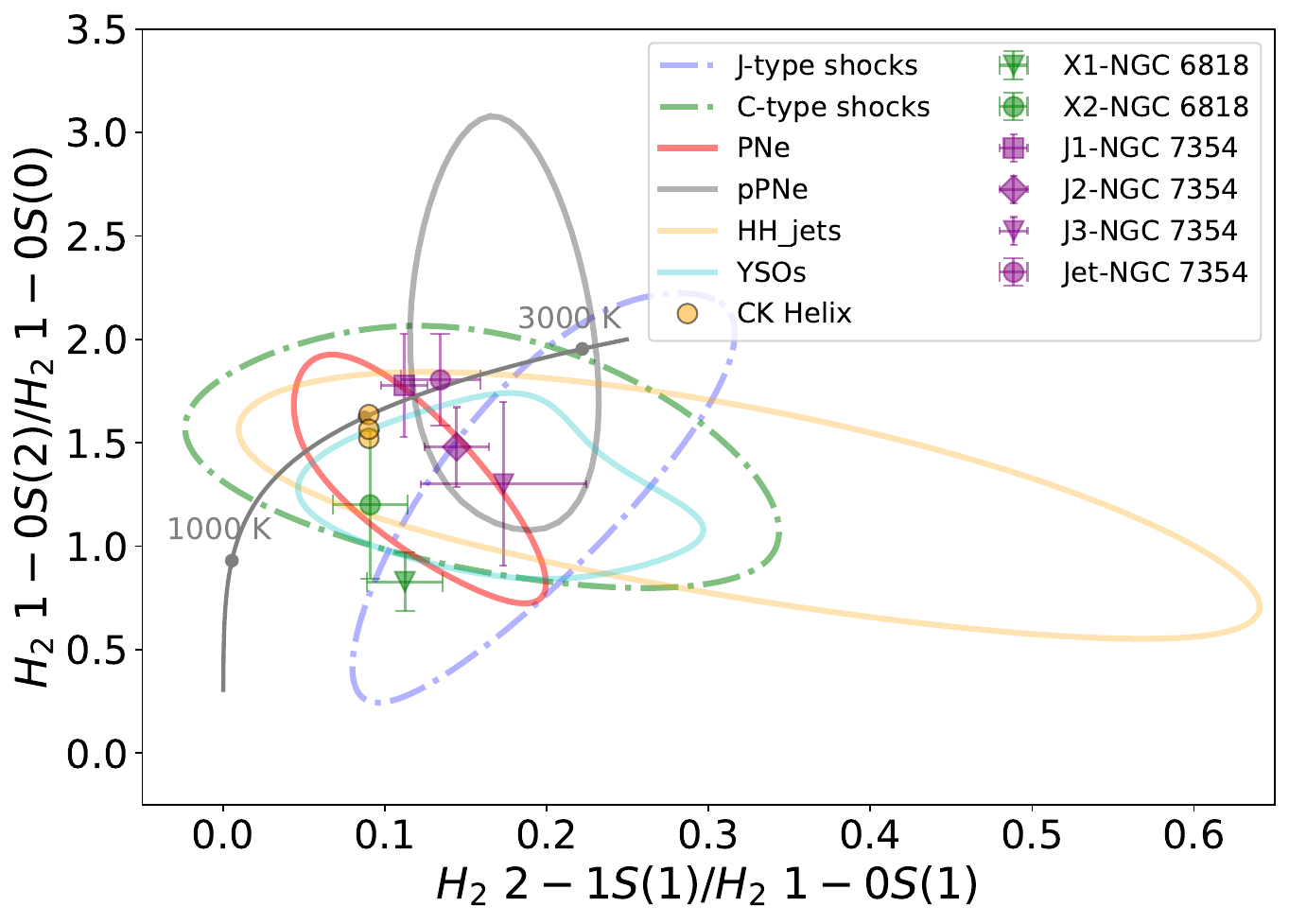}
       \caption{\textit{Left panel}: H$_2$~2--1S(1)/1--0S(1) versus H$_2$~1--0S(2)/1--0S(0) DD showing the LTE  thermal curve at 1000-3000~K. Also shown are models and observational data gathered by \citet{1994ApJ...427..777M}. Thermal UV excitation models \citep[][]{1989ApJ...338..197S} are represented by open circles; X-ray heating models \citet{1983ApJ...269..560L,1990ApJ...363..464D} by open triangles; non-thermal models \citep[][]{1987ApJ...322..412B} by crosses; and a shock-heating model \citep[][]{1989MNRAS.236..929B} by a filled circle. Observational data of SNRs \citep[][]{1991MNRAS.253..662M,1990A&A...240..453O} are shown as filled circles, and for PDRs \citep[][]{1991A&A...250..475I,1989ApJ...336..207T} as filled triangles. \textit{Right panel}: The same DD with observational data for PNe \citep{1993MNRAS.263..695R,1999ApJS..124..195H,2006AJ....131.1515L,2008AJ....135.1441R,2015MNRAS.453.1888M} outlined by a solid red line; pPNe by a solid gray line \citep{2005ApJ...629.1040K,2015MNRAS.447.1080G}; HHs by a solid orange line \citep{1999MNRAS.308..539D,2002A&A...393.1035N,2005A&A...441..159N,2006ApJ...641..357T,2010A&A...511A...5G,2011A&A...528A...3D}; and YSOs a solid light blue line \citep{2015A&A...573A..82C}. Also shown are data for a CK in Helix presented as orange circles \citep{2007MNRAS.382.1447M}, and contours from C-type (green dash-dotted line) and J-type shock (blue dash-dotted line) models \citep{2023A&A...675A..86K}. See text for more details. }
       \label{fig:H2_H2}%
\end{figure*}

To further investigate whether the H$_2$ emission in the analyzed structures is consistent with LTE conditions inferred from the $T_{\rm vib}~{\rm vs.}~T_{\rm rot}$ comparison, we now examine a classical H$_2$ DD that directly contrasts thermal and non-thermal excitation regimes.
Figure~\ref{fig:H2_H2} (left panel) shows the H$_2$~2--1S(1)/1--0S(1) versus H$_2$~1--0S(2)/1--0S(0) DD for the different structures in our sample, together with models and observational data gathered by \citet{1994ApJ...427..777M}. These include thermal UV excitation models \citep[][]{1989ApJ...338..197S}, X-ray heating models \citet{1983ApJ...269..560L,1990ApJ...363..464D}, a shock-heating model \citep[][]{1989MNRAS.236..929B}, as well as observations of supernova remnants (SNRs) \citep[][]{1991MNRAS.253..662M,1990A&A...240..453O} and photodissociation regions (PDRs) \citep[][]{1991A&A...250..475I,1989ApJ...336..207T}. 

The diagram also includes the expected locations of thermal emission at 1000-3000~K, computed under the LTE approximation using the Boltzmann equation, the Einstein A coefficient from \citet{1977ApJS...35..281T}, and the energy levels from \citet{1993MNRAS.263..695R}. Points located close to this thermal curve are consistent with LTE conditions and predominantly thermal collisional excitation. In contrast, significant deviations from this curve --~particularly toward the lower-right region of the diagram~-- are indicative of non-thermal excitation mechanisms, such as UV fluorescence. As noted by \citet{1994ApJ...427..777M}, the LTE thermal curve lies well separated from the non-thermal model points, which populate the lower right corner.

The right panel of Fig.~\ref{fig:H2_H2} displays the same DD compared with contour regions for PNe \citep{1993MNRAS.263..695R,1999ApJS..124..195H,2006AJ....131.1515L,2008AJ....135.1441R,2015MNRAS.453.1888M}, proto-PNe (pPNe) \citep{2005ApJ...629.1040K,2015MNRAS.447.1080G}, Herbig–Haro objects (HHs) \citep{1999MNRAS.308..539D,2002A&A...393.1035N,2005A&A...441..159N,2006ApJ...641..357T,2010A&A...511A...5G,2011A&A...528A...3D}, and young stellar objects (YSOs) \citep{2015A&A...573A..82C}. Also shown are three regions within a cometary knot (CK) in the Helix nebula reported by \citet{2007MNRAS.382.1447M}, for which C-type shocks were found to reproduce the observed H$_2$ emission under plausible local conditions.
Therefore, we further include contours corresponding to C- and J-type shock models from \citep{2023A&A...675A..86K} considering (i) n$_H$ between 10$^{4}$ and 10$^{8}$ to be consistent with models from \citet{2020ApJ...889...13B}, 
(ii) velocities from 15~km~s$^{-1}$ to 60~km~s$^{-1}$, consistent with velocities found for these structures \citep[][]{2003A&A...400..161B,2010AJ....139.1426C}, and the dimensionless magnetic field scaling factor ($b = B/\sqrt{n_{\mathrm{H}}}$) between 0.1 and 1.0, as sampled in the shock model grid of \citet{2023A&A...675A..86K}.

In this diagram, only the whole jet (Jet) and its J1 segment (the closest to the CSPN) in NGC~7354 lie close to the thermal excitation curve corresponding to temperatures of approximately 2000–3000~K. It can be noted that our data point overlap with those of YSOs, a subset of HHs, and other sources commonly associated with collisionally excited H$_2$ emission, while only the structures associated with NGC~7354 overlap with the proto-PNe regime. Moreover, the LISs studied in this work lie also within the region occupied by C-type shock models, as concluded by \citet{2007MNRAS.382.1447M} through the study of a CK in the Helix nebula. They also intersect with the regime of PNe with detected H$_2$ emission, suggesting that the LISs share similar excitation conditions with these nebulae. In contrast, none of the analysed structures occupy the region of the diagram typically associated with non-thermal UV fluorescence.  This picture is fully consistent with the T$_{\rm vib}$–T$_{\rm rot}$ comparison and the diagrams discussed above, reinforcing the conclusion that the H$_2$ emission arises under thermal conditions and is close to LTE.

Finally, it is important to note that thermalized H$_2$ emission can result not only from shocks but also from X-ray heating or UV irradiation in dense environment. Therefore, while this diagram strongly supports a primarily thermal excitation regime, it does not uniquely constrain the underlying energy source.

\subsection{[Fe~{\sc ii}] as diagnostic of atomic gas heating}

While the H$_2$-based DDs provide valuable insight into the excitation conditions of the molecular gas, they do not directly probe the mechanisms responsible for heating and ionizing the atomic gas. 
Recently, \cite{2024A&A...689A..70A} reported the first detection of the [Fe~{\sc ii}] 1.644~$\mu$m emission lines originating from several LISs in four PNe, based on narrowband imaging, while \citet{2025MNRAS.541.1628B} detected [Fe~{\sc ii}] $\lambda$8617 emission from LISs and clumps using MUSE. To further constrain the nature of the processes governing the LISs, we now examine the behaviour of the [Fe~{\sc ii}] emission through longslit spectroscopy, which traces the partially ionized gas and is particularly sensitive to shocks and X-ray irradiation.

[Fe~{\sc ii}] is a well-established tracer of shock excitation, with radiative atomic shocks observed in the near- and mid-IR usually show prominent forbidden lines from Fe$^{+}$ \citep{2016JKAS...49..109K}. This behaviour arises from a number of unique properties that make [Fe~{\sc ii}] emission particularly strong \citep[][]{1989ApJ...342..306H,1989A&A...214..307O}.
On the one hand, Fe$^{+}$ has several low-excitation energy levels, that can be readily excited in shocked gas, and whose transitions produce numerous emission lines, particularly in the near-IR \citep[][]{1989ApJ...342..306H}. On the other hand, the ionization potential of a neutral iron atom is only 7.9~eV\footnote{From \href{https://physics.nist.gov/PhysRefData/ASD/ionEnergy.html}{NIST Database}}, allowing an excited region to develop behind the shock front, where Fe is ionized to Fe$^{+}$ by FUV photons produced at the shock front, while hydrogen atoms remain mostly neutral. This contrasts with photoionized regions, where iron atoms are predominantly found in higher ionization states, unless the ionizing radiation is sufficiently hard to penetrate deeply into the interstellar cloud, as in the case of active galactic nuclei (AGN). Finally, the Fe abundance can be significantly enhanced in shocked regions due to dust grain destruction, since shocks are efficient at destroying the grains in which iron is otherwise locked \citep[with $\sim$99\% of the iron in the ISM being depleted onto dust grains][]{1989A&A...214..307O}.

\begin{figure}[ht]
       \centering
       \includegraphics[width=0.98\columnwidth]{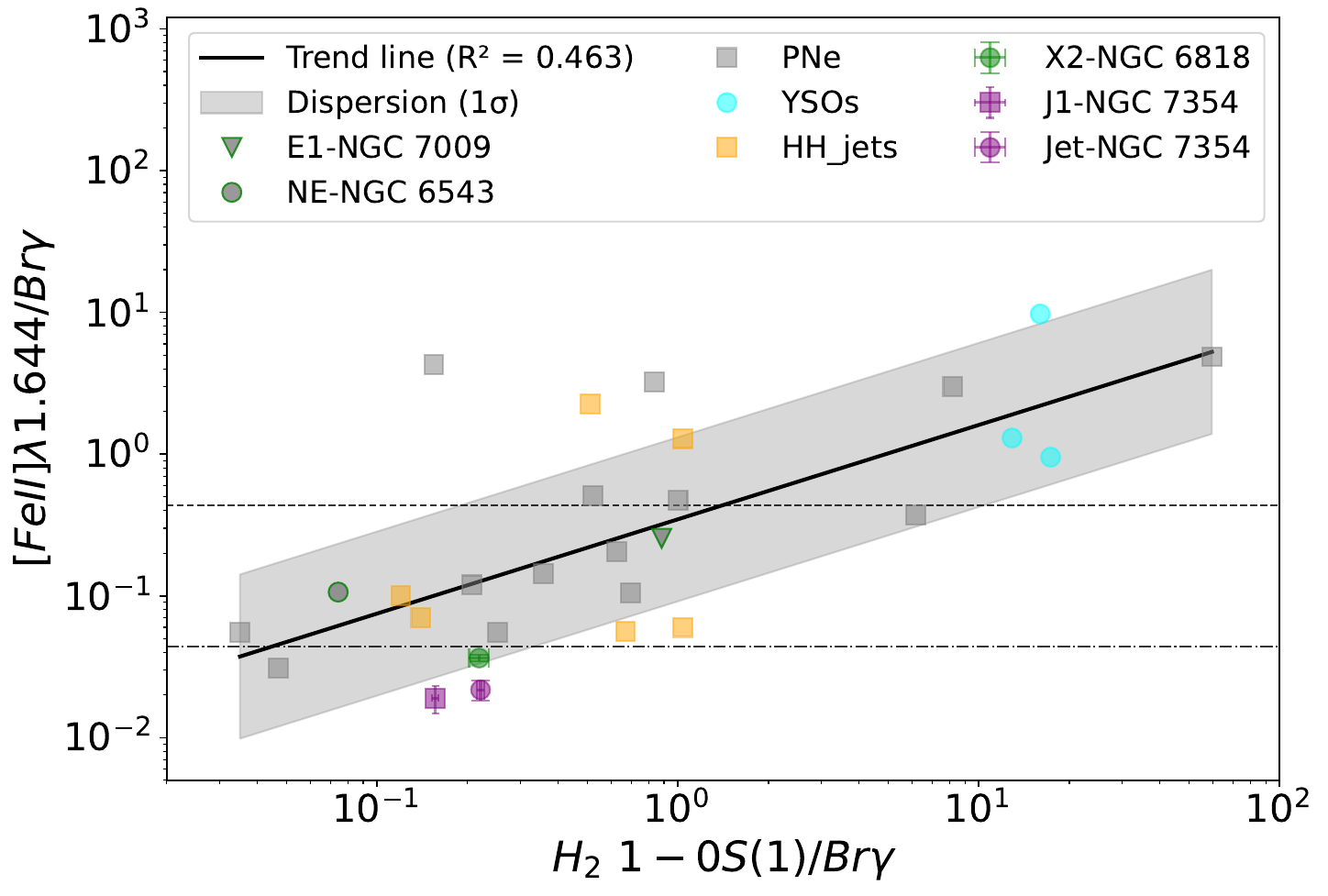}   
       \caption{H$_2$ 1–0 S(1)/Br$\gamma$ versus [Fe \textsc{ii}] 1.644 $\mu$m/Br$\gamma$ diagram for the LISs analyzed in this work, together with two additional LISs from the literature \citep[E1 in NGC 7009 and NE in NGC 6543;][]{2024A&A...689A..70A,2020MNRAS.493.3800A}. Also shown are comparison samples of PNe, YSOs, and HH objects compiled from the same references as in Fig.~\ref{fig:H2_H2}. The horizontal dashed line at [Fe \textsc{ii}]/Br$\gamma$ = 0.437 marks the lower limit above which SNRs are typically found, while the dash-dotted line at [Fe \textsc{ii}]/Br$\gamma$ = 0.0437 indicates the lower limit expected for H \textsc{ii} regions, corresponding to shock-dominated and photoionization-dominated excitation, respectively \citep{2015PKAS...30..145K}.}
       \label{fig:ferrum}%
\end{figure}
\vspace{-0.1cm} 

As a consequence, [Fe~{\sc ii}] emission lines arising from shocked gas are considerably stronger than those from photoionized regions; for instance, [Fe~{\sc ii}]1.257/Pa$\beta$=1–10 for Seyfert galaxies \citep[][]{2000ApJ...528..186M} or $>$0.1 for SNRs compared to $\sim$0.01-0.03 for Orion H{\sc ii} region \citep[][]{2015PKAS...30..145K}. Assuming [Fe~{\sc ii}]1.644/Br$\gamma$ = 4.37 $\cdot$ [Fe~{\sc ii}]1.257/Pa$\beta$ \citep[][]{1989A&A...214..307O}, this translates into expected values of [Fe~{\sc ii}]~1.644/Br$\gamma \gtrsim$ 0.437 for shock-dominated regions (such as SNRs), while photoionized regions are expected to exhibit [Fe~{\sc ii}]1.644/Br$\gamma \gtrsim 0.0437$, an order of magnitude lower.

Figure~\ref{fig:ferrum} shows the H$_2$~1--0S(1)/Br$\gamma$ versus [Fe~{\sc ii}]~1.644$\mu$m/Br$\gamma$ diagram for the structures analyzed in this work, together with literature data for PNe, YSOs, and HH objects. A clear positive correlation is observed, with a moderate coefficient of determination ($R^{2} \sim 0.463$). Although this correlation does not necessarily imply that H$_2$ and [Fe~{\sc ii}] originate from the same volume of gas or are excited by the same physical mechanism, it supports a scenario in which the molecular and atomic gas phases are physically coupled within LISs. The LISs analyzed here exhibit integrated [Fe~{\sc ii}]~1.644~$\mu$m/Br$\gamma$ ratios below 0.0437, with values of 0.036 for X2 in NGC~6818, 0.019 and 0.022 for J1 and Jet of NGC~7354, respectively. These values favour a photoionization-dominated contribution to the atomic gas excitation. Interestingly, \citet{2024A&A...689A..70A} have reported values ranging from 0.15 up to 7, covering a regime from photo-dominated and shock-dominated LISs.

A similar trend between H$_2$ and [Fe~{\sc ii}], was also reported by \citet{2004A&A...425..457R} for AGNs, suggesting that both sets of lines may be excited by a common mechanism. In particular, X-ray radiation has been proposed as a viable process, since it can simultaneously produce strong [Fe~{\sc ii}] emission through dust destruction and yet not destroy the H$_2$ molecules. 
However, according to \citet{2002MNRAS.331..154R}, if X-rays play a significant role in the excitation of molecular gas, very large [Fe~{\sc ii}]/Br$\gamma$ ratios are expected, typically of the order of $\sim$20. The significantly lower values measured in our LISs argue against X-ray-dominated excitation, despite the observed correlation between H$_2$ and [Fe~{\sc ii}].

Among the PNe compiled from the literature \citep{1999ApJS..124..195H}, six objects lie above the [Fe~{\sc ii}]/Br$\gamma = 0.437$ threshold. Most of these correspond to large-scale structures rather than LISs in the classical sense: Hb~12 (at 3.7''~E and 3.7''E~2''~S), M~1-92 ("NW") and AFGL~618 (at 2.4''~E) are lobes or regions offset from the central star. Two exceptions are NGC~2440 ("NE", a clump) and M~2-9 ("Nknot"), which are morphologically closer to LISs and may genuinely exhibit higher [Fe~{\sc ii}] emission, possibly reflecting differences in physical conditions or excitation mechanisms. Nevertheless, \citet{2020MNRAS.497.3180A} showed that the H$_2$~1--0~S(1)/Br$\gamma$ ratio is sensitive to the slit configuration: when the slit covers only the region around the H$_2$ emission peak, Br$\gamma$ is minimized and the ratio can reach artificially high values. 
More broadly, any ratio combining lines that originate in spatially distinct regions, will be sensitive to the slit configuration.

The low [Fe~{\sc ii}]/Br$\gamma$ ratios measured in the LISs, combined with the evidence for thermally excited H$_2$, argue against excitation scenario dominated by strong shocks or X-ray irradiation. Instead, the observations are consistent with a picture in which the atomic gas is primarily photoionized by the CSPNe, while the molecular gas is thermally excited in dense, UV-irradiated regions, possibly with a minor contribution from weak shocks that do not significantly affect the observed line ratios, but can still lead to the destruction of dust grains and the release of Fe into the gas phase.

\section{Spatial distribution of emission lines}\label{sect_spatial}

\begin{figure*}[ht]
       \centering
       \includegraphics[width=0.98\columnwidth]{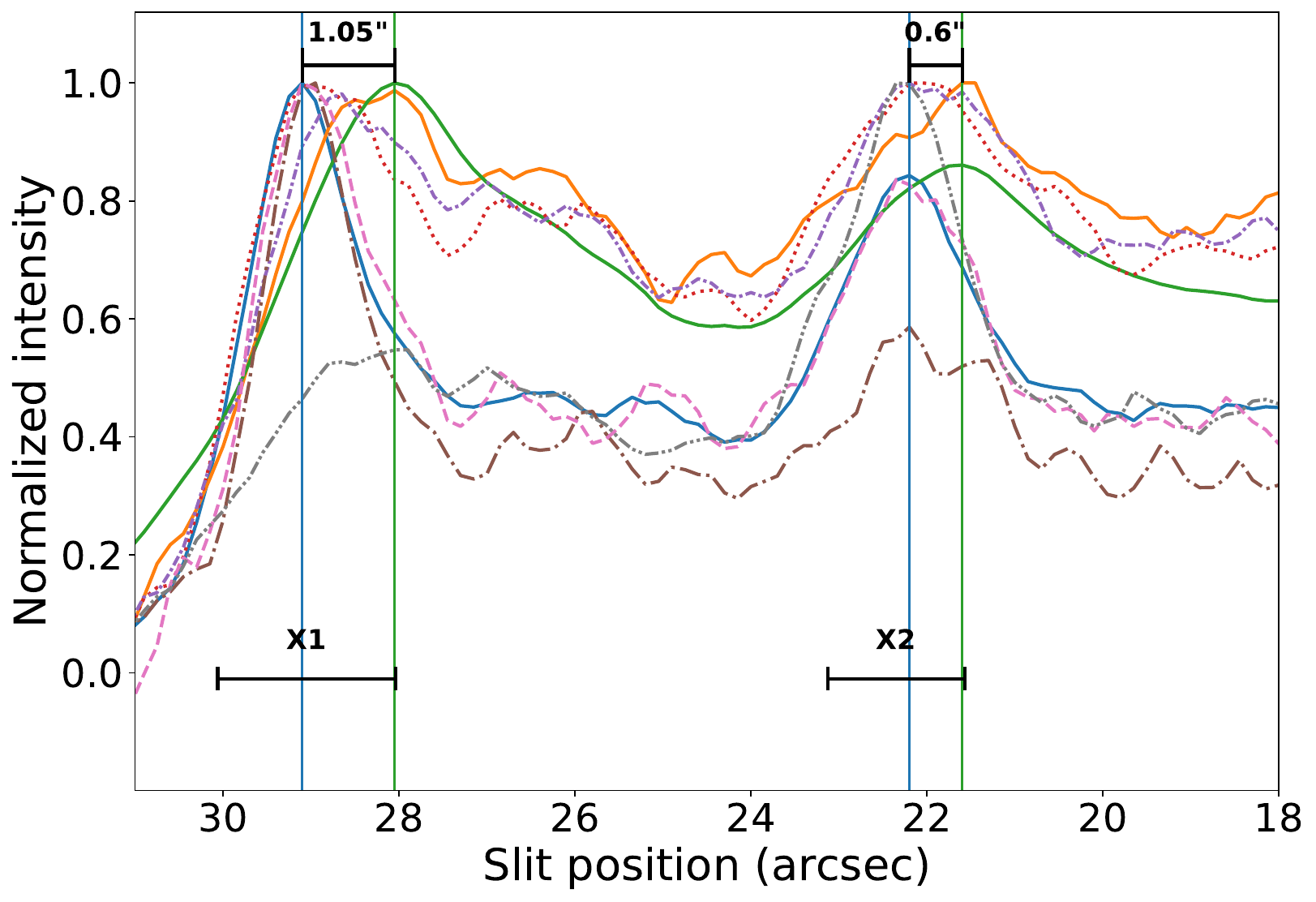}
       \includegraphics[width=0.98\columnwidth]{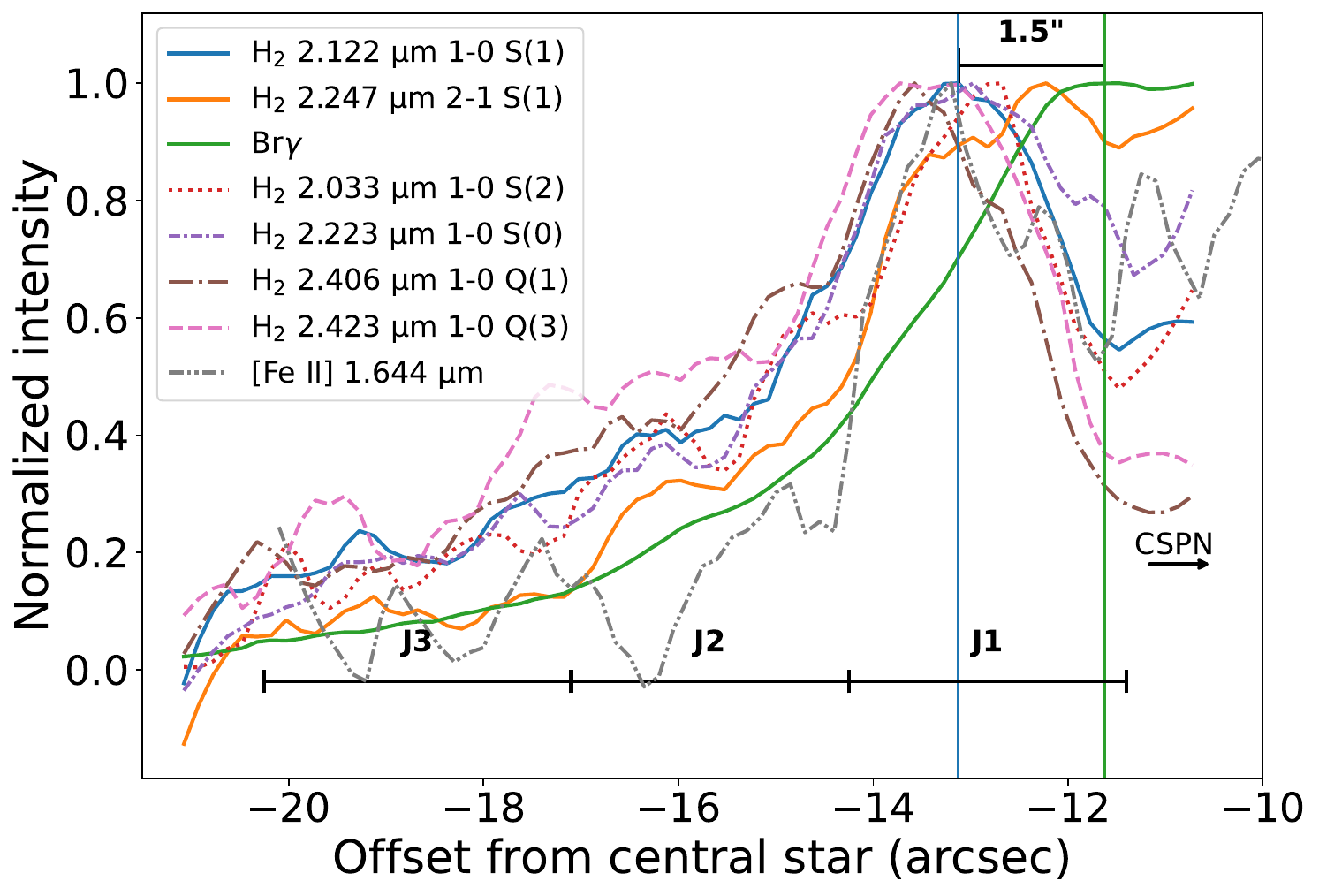}\\
       \caption{Spatial profiles of H$_2$, Br$\gamma$, and [Fe~{\sc ii}] 
    1.644~$\mu$m emission lines along the slit for NGC~6818 (\textit{left}) and 
    NGC~7354 (\textit{right}). All profiles are normalized to their peak value. 
    Vertical solid lines mark the positions of the peaks identified 
    in each LIS (X1, X2 for NGC~6818; Jet for NGC~7354) for H$_2$~1--0~S(1) and Br$\gamma$.}
       \label{fig:spatial_profiles}%
\end{figure*}

Figure~\ref{fig:spatial_profiles} shows the normalized spatial profiles of the H$_2$~1--0~S(1) (2.122~$\mu$m), H$_2$~2--1~S(1) (2.247~$\mu$m), 
Br$\gamma$ (2.166~$\mu$m), and [Fe~{\sc ii}] (1.644~$\mu$m) emission lines along the slit, for both NGC~6818 and NGC~7354. These profiles were obtained directly from the two-dimensional spectra, without continuum subtraction, to trace the spatial variation of the emission along the slit. 
Two distinct spatial peaks are identified in NGC~6818, labelled X1 and X2 (corresponding to its two LISs). In NGC~7354, we define three spatial segments along the jet (J1–J3). These regions are consistent with those defined in Section~\ref{sect3}.
Among them, J1 is the innermost (closest to the central star) and has the highest signal-to-noise ratio, whereas J3 is the outermost and the noisiest segment.

In both objects, the Br$\gamma$ emission peaks at a systematically different 
spatial position with respect to the H$_2$ 1--0\,S(1) line, result that was also found in \citet{2020MNRAS.493.3800A}. 
The observed stratification is therefore consistent with the classical geometry of a PDR: Br$\gamma$ peaking near the source of UV radiation (CSPN) representing the external layer of LISs that surround the internal H$_2$ emission. 
In NGC~6818, the spatial separation between the Br$\gamma$ and H$_2$~1--0~S(1) peaks amounts to $\sim$1.05\arcsec\ at X1 and $\sim$0.6\arcsec\ at X2 (visually the closest to the central star); in NGC~7354, a separation of $\sim$1.5\arcsec\ is measured along the jet direction, consistent with results derived independently from narrow-band imaging \citep[offset of $\sim$1.2\arcsec\,][]{2024A&A...689A..14A}. 
Notably, the [Fe~{\sc ii}] 1.644~$\mu$m emission peaks at the same spatial  position as the H$_2$~1--0~S(1) in both objects, suggesting that the partially ionized atomic and warm molecular gas are co-spatial within the LISs, and may be excited by the same physical process, possibly FUV photons generated at the shock front, or their spatial separation may simply be smaller than the spatial resolution of our data.

Regarding the H$_2$~2--1~S(1) (2.247~$\mu$m) line, a spatial offset with respect to H$_2$~1--0~S(1) is apparent in the profiles of both objects. If real, this would suggest a physically interesting result: the 2--1~S(1) transition arises from the $v=2$ vibrational level, which requires 
significantly higher excitation energies than the $v=1\to0$ transitions (all of them peaking at the same position as 2.122~$\mu$m), and a spatial displacement between these two lines would constitute direct evidence for vibrational excitation stratification within the molecular gas \citep[][]{2007RPPh...70.1099O}. However, since the profiles presented here were not continuum-subtracted, and since the H$_2$ 2--1\,S(1) line is relatively faint, we cannot rule out that the apparent offset is at least partially induced by contamination from the underlying continuum emission. A definitive assessment of the spatial distribution of the 2--1~S(1) line across the two‑dimensional spectrum would require a detection with a higher signal‑to‑noise ratio. Future JWST data may provide a more conclusing answer to that. We therefore refrain from drawing strong conclusions regarding this offset, and focus the discussion on the robust Br$\gamma$--H$_2$ 1--0\,S(1) and [Fe~{\sc ii}]--H$_2$ 1--0\,S(1) spatial stratification.

The spatial stratification between Br$\gamma$, H$_2$ 1--0\,S(1), and [Fe~{\sc ii}], is consistent with a photoevaporation scenario within the LISs proposed by \citet{2024A&A...689A..70A}, in which Br$\gamma$ traces the photoionized gas closer to the CSPN, while H$_2$ and [Fe~{\sc ii}] arise farther out. 

Figure~\ref{fig:spatial_profiles2} show the variation of the H$_2$~2.122/Br$\gamma$, H$_2$~2.247/Br$\gamma$ and H$_2$~2.122/2.247 line ratios along the slit. These ratios are normalized to the maximum value of each profile, highligthing their relative variations and emphasizing the structures of interest. The H$_2$~2.122/Br$\gamma$ and H$_2$~2.122/2.247 line ratios exhibit clear peaks at the positions of the LISs in NGC~6818 while the surrounding nebula is characterized by significantly lower values. In the case of the Jet in NGC~7354, both line ratios are very low close to the central star and increase with distance along the jet toward its outer regions. This increase can be attributed to the more rapid decline of the Br$\gamma$ emission along the jet, as well as to the significant contribution of the lower-excitation \(v=1\rightarrow0\) H$_2$ emission relative to the \(v=2\rightarrow1\) transitions. In contrast to these line ratios, the H$_2$~2.247/Br$\gamma$ remains relatively uniform along the slit, showing only modest spatial variations, as the H$_2$~2.247 line is weak.

\section{Discussion}\label{discussion}

This work presents near-IR longslit spectra of the LISs in two PNe with different morphologies and kinematic properties: NGC~6818 and NGC~7354.

NGC~6818 is an elliptical PN hosting a pair of LISs (X1 and X2) oriented along PA~=~130$^{\circ}$. These LISs are commonly referred to as "mustaches" in the literature and expand at $\sim 22~km~s ^{-1}$ \citep{2003A&A...400..161B}, well within the velocity range covered by the C-type shock models included in Fig.~\ref{fig:H2_H2}. 
In contrast, NGC~7354 hosts a collimated jet oriented at PA~=~24.4$^{\circ}$, with launch velocity of $\sim 60~km~s ^{-1}$\citep{2010AJ....139.1426C}. The currently measured velocity is lower than the predicted ones, which has been interpreted as the result of deceleration through interaction with the polar caps \citep{2010AJ....139.1426C}. In our data, the jet is resolved into three spatial segments (J1, J2 and J3), with the complete jet (Jet) showing the highest excitation temperature (T$_{ex}\sim$2450~K) among all the analyzed structures. 

\subsection{Excitation conditions of the molecular gas}
The H$_2$ excitation diagrams, the T$_{\rm vib}$--T$_{\rm rot}$ comparison, and the classical H$_2$ diagnostic diagrams all point consistently to the same conclusion: the molecular hydrogen emission in the LISs of both PNe arises under predominantly thermal conditions, close to LTE. 
The excitation temperatures derived from the diagrams (T$_{\rm ex}\sim$1860--2450~K) are in reasonable agreement with the vibrational temperatures estimated independently from line ratios, and the structures lie well within the region occupied by C-type shock models and other sources associated with collisionally excited H$_2$ emission, such as YSOs and a subset of HH objects. C-type shocks, in which magnetic fields gradually compress and heat the gas without a sharp discontinuity in density or temperature \citep{2004ApJ...614L..69H}, are broadly consistent with the moderate velocities measured in both NGC~6818 ($\sim$22~km~s$^{-1}$; \citealt{2003A&A...400..161B}) and NGC~7354 ($\sim$60~km~s$^{-1}$ at launch, currently decelerated; \citealt{2010AJ....139.1426C}). In contrast, J-type shocks, which produce an abrupt discontinuity at the shock front \citep{2004ApJ...614L..69H}, would be expected to produce stronger [Fe~{\sc ii}] emission and higher H$_2$ vibrational ratios than observed here.
However, the systematic offset between T$_{\rm vib}$ and T$_{\rm rot}$ --~with T$_{\rm rot}$ consistently lower than T$_{\rm vib}$ across all structures~-- prevents a straightforward interpretation. As discussed in Sect.~\ref{sect:h2diag}, this behavior can arise either from a genuine secondary contribution of non-thermal excitation (e.g., UV fluorescence), or from the thermalization of UV-excited H$_2$ in a dense gas, where collisional de-excitation drives the level populations toward LTE. In either case, the available diagnostics do not allow us to uniquely identify the underlying energy source, since shocks, UV irradiation, and X-ray heating can all produce thermalized H$_2$ emission under appropriate conditions. The detection of levels arising only from v~=~1 $\rightarrow$ 0 and v~=~2 $\rightarrow$ 1 transitions further limits our ability to identify the non-thermal signature that would emerge at higher vibrational levels.

\subsection{Excitation conditions of the atomic gas}
The [Fe~{\sc ii}]/Br$\gamma$ ratios measured in the LISs analyzed here are well below the threshold expected for shock-dominated regions, and fall instead in the range consistent with photoionization. Combined with the positive correlation observed between H$_2$ and [Fe~{\sc ii}] emission (Fig.~\ref{fig:ferrum}), this suggests that the molecular and atomic gas phases are physically coupled within the LISs, but the dominant excitation of the atomic gas is not driven by strong shocks. The very large [Fe~{\sc ii}]/Br$\gamma$ ratios expected from X-ray-dominated excitation \citep{2002MNRAS.331..154R} are also not observed, arguing against a significant X-ray contribution. This is further supported by \citet{1996ApJ...466..561M}, who showed that X-ray-dominated regions produce [Fe~{\sc ii}] intensities significantly stronger than those observed here, and that the H$_2$ emission in such regions arises primarily from thermal excitation in the warm atomic zone.
Taken together, the low [Fe~{\sc ii}]/Br$\gamma$ ratios and the thermally excited H$_2$ are most naturally explained by a scenario in which the LISs are dense clumps being photoionized from the outside by the radiation field of the CSPN, with a possible minor contribution from slow-moving internal shocks that are not energetic enough to significantly alter the observed line ratios.

\subsection{Spatial stratification and the photoevaporation scenario}
The spatial stratification observed in both objects --~Br$\gamma$ peaking closer to the CSPN, H$_2$~1--0\,S(1) and [Fe~{\sc ii}] co-spatial further out~-- is the most direct observational evidence in support of a photoevaporation scenario. This geometry is precisely what is expected when a dense clump is exposed to the ionizing radiation of the central star: the outer layers are progressively ionized (traced by Br$\gamma$), while the inner molecular core remains self-shielded and is traced by H$_2$ emission. The co-spatiality of [Fe~{\sc ii}] with H$_2$ suggests that iron is being released into the gas phase by a slow shock at the interface between the ionized and molecular regions, consistent with the interpretation proposed by \citet{2024A&A...689A..70A} for LISs in five other PNe.

The smaller Br$\gamma$--H$_2$ offset observed at X2 in NGC~6818 with respect to X1 is also worth noting. X2 is the structure \textit{visually} closest to the CSPN, and according to the photoevaporation model of \citet{2018ApJ...861...45M} as cited in \citet{2024A&A...689A..14A}, the offset between H$_2$~1--0\,S(1) and Br$\gamma$ becomes smaller when the flux of ionizing photons increases or the illuminated clump is smaller. Both conditions may apply to X2, making this a qualitatively consistent result.

\subsection{Molecular and ionized masses}

\begin{figure}[ht]
       \centering
       \includegraphics[width=0.98\columnwidth]{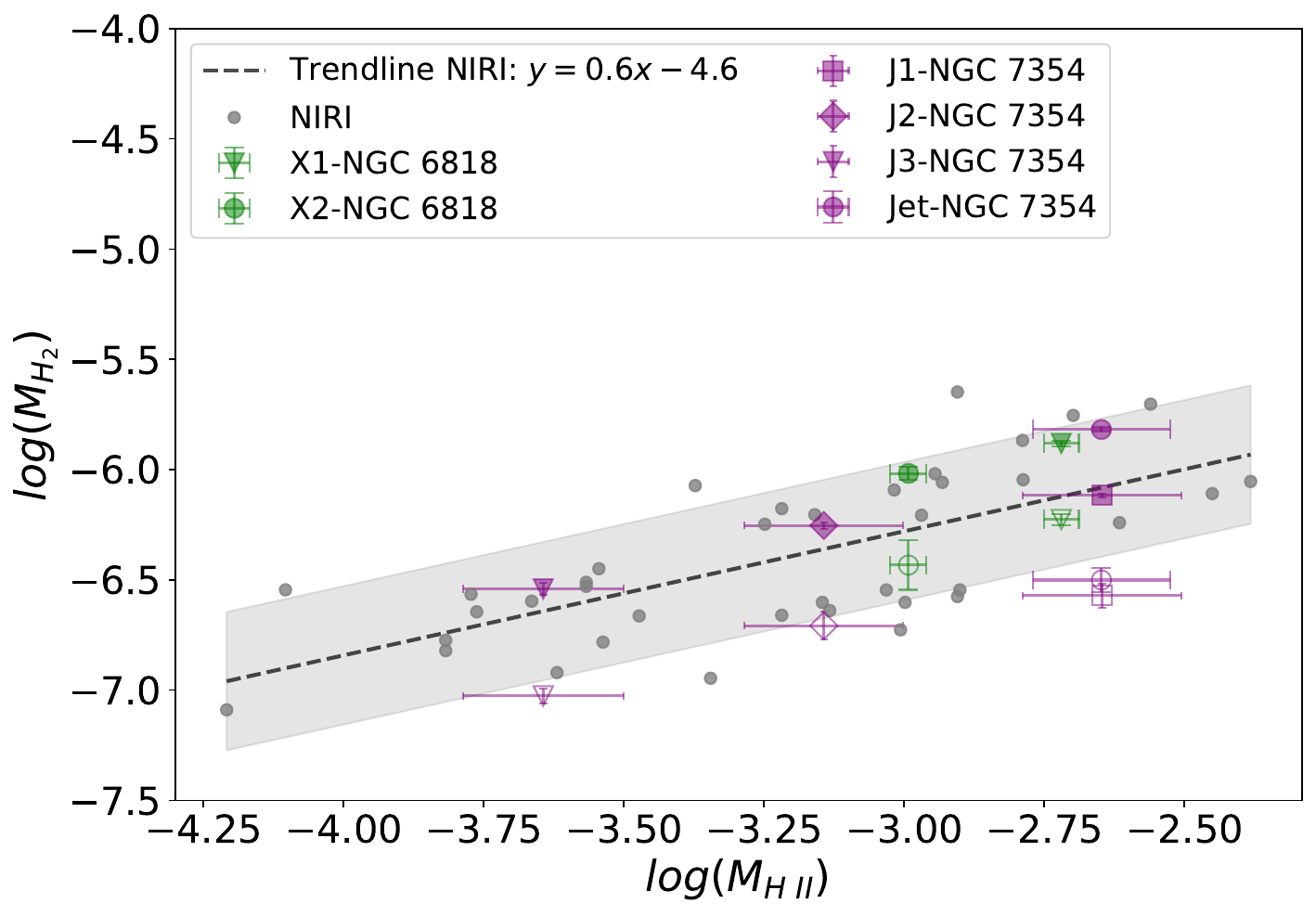}
       \caption{log(M$_{\rm H_2}$) versus log(M$_{\rm H~II}$) for the LISs analyzed in this work. Filled symbols correspond to warm molecular masses derived from H$_2$~1--0\,S(1) line emission, while open symbols correspond to masses estimated from H$_2$ column densities obtained from the excitation diagrams. For each structure, the filled and open symbols share the same M$_{\rm ion}$. Gray points show NIRI measurements from \citet{2017MNRAS.465.1289A,2020MNRAS.493.3800A} with the dashed line showing the trend derived by \citet{2025A&A...704A..59M}.}
       \label{fig:masses}%
\end{figure}

Figure~\ref{fig:masses} shows the warm molecular masses derived for the LISs in this work from the H$_2$~1--0\,S(1) emission lines, plotted against the ionized masses estimated from Br$\gamma$, following the methodology of \citet{2025A&A...704A..59M}. The new data points from NGC~6818 and NGC~7354 are consistent with the trend previously established from narrow-band imaging data \citep{2017MNRAS.465.1289A,2020MNRAS.493.3800A}, with ionized masses systematically larger than the warm molecular masses by factors of $\sim$800--3000, within the broader range of $\sim$200--5000 reported by \citet{2025A&A...704A..59M}.

Also shown are the warm molecular masses estimated independently from the H$_2$ column densities (N(H$_2$)) obtained through the excitation diagrams (M$_{\rm warm~column}$, open symbols, where the ionized mass used is the same as for the filled symbols in each case). Both estimates agree within a factor of $\sim$2.2--4.8 (see Table~\ref{tab:masses}), confirming that the two independent methods yield consistent results. 
Regarding the excitation temperatures along the jet in NGC~7354, an interesting spatial trend is apparent in Table~\ref{table_TexNH2}: T$_{\rm rot}$ decreases with distance from the central star (from $\sim$2010~K in J1 to $\sim$1280~K in J3), while T$_{\rm vib}$ shows the opposite behavior, increasing outward. This divergence between the two temperature estimators suggests that the relative contribution of non-thermal excitation becomes more pronounced in the outer jet segments, further from the ionizing source.

\section{Conclusions}\label{conclusions}
We have presented a near-IR longslit spectroscopic study of low-ionization structures in planetary nebulae, obtained with GNIRS at Gemini North for NGC~6818 and NGC~7354. The main results can be summarized as follows:
\begin{itemize}
\item H$_2$, Br$\gamma$, and [Fe~{\sc ii}] emission are detected in almost all the structures studied, demonstrating that LISs host simultaneously ionized, atomic, and warm molecular gas phases.
\item The H$_2$ excitation diagrams and the comparison between vibrational and rotational temperatures indicate that the molecular gas is excited under predominantly thermal, near-LTE conditions, with excitation temperatures in the range $\sim$1860--2450~K, consistent with the high densities (n$_{\rm total}\sim 10^{4}-10^{5}$; Appendix~\ref{appendixB}) inferred independently from the gas-to-extinction relation. A secondary non-thermal contribution cannot be excluded, but is not required to explain the observations.
\item The [Fe~{\sc ii}]/Br$\gamma$ ratios measured in the LISs are well below the values expected for shock-dominated or X-ray-dominated regions, and are instead consistent with photoionization as the dominant excitation mechanism.
\item A clear spatial stratification is observed, with Br$\gamma$ peaking closer to the central star and H$_2$~1--0\,S(1) and [Fe~{\sc ii}] co-spatial at larger distances. This is consistent with photoevaporation, where dense LIS clumps are progressively ionized from the outside by the CSPN radiation field.
\item The warm molecular masses derived from H$_2$ line emissions are consistent with those estimated independently from the excitation diagrams, and the ionized masses remain systematically larger by factors of hundreds to thousands, extending the trend established by \citet{2025A&A...704A..59M} to a larger sample.
\end{itemize}

These results call for two complementary follow-up directions. On one hand, expanding the sample of LISs observed in the H and K bands with longslit or integral-field spectroscopy would improve the statistical significance of the excitation and mass trends identified here. 
On the other hand, observations at longer wavelengths (L band, $\sim$3.5~$\mu$m) would access higher vibrational transitions of H$_2$ (v$\geq$3), where the non-thermal signature of UV fluorescence becomes detectable and would allow a more definitive discrimination between the excitation mechanisms at play.

\begin{acknowledgements}
We thank the anonymous referee for her/his comments and suggestions, which helped improve the manuscript.
This research project was supported by the Consejo Nacional de Investigaciones Científicas y Técnicas (CONICET, Argentina).
Based on data obtained from the International Gemini Observatory (managed by AURA for NSF's NOIRLab) on behalf of the Gemini partnership. The Brazilian NGO has awarded a fraction of the observational time for this project (2.8 hours, GN-2024A-018). 
MBM and CGD acknowledge the support of the ``Agencia Nacional de promoci\'on de la investigaci\'on, el desarrollo tecnol\'ogico y la innovaci\'on'' through project PICT-2021-GRF-TII-00442, and the support of CONICET through project PIP 11220210100520CO.
MBM and CGD acknowledge the support of the ``Secretar\'ia de Ciencia y Tecnolog\'ia'' of the ``Universidad Nacional de C\'ordoba'' through grant SeCyT "Consolidar" 2023 NRO:33620230100103CB. DRG acknowledges the grants of the Brazilian agencies FAPERJ (E-26/200.627/2023 and E-26/200.467/2026) and CNPq (315307/2023-4). 

\end{acknowledgements}

%
   \bibliographystyle{aa} 
   \bibliography{refes} 

\begin{appendix}
\onecolumn
\section{Line flux measurements}
\label{appendixA}


\begin{table*}[ht!]
\caption{Emission line fluxes ($\times 10^{-15}$ erg s$^{-1}$ cm$^{-2}$) in the LISs of NGC~6818. H-band lines are above the dashed line, K-band lines below. Line identifications are primarily based on the \href{https://linelist.pa.uky.edu/atomic/}{Atomic Line List} (air wavelengths), while H$_2$ transitions follow \citet{2011A&A...528A..74A} and \citet{1987ApJ...322..412B}.}
\centering
\small
\begin{tabular}{l l c c c c c}
\hline
$\lambda_{\rm rest}$ (\AA) & $\lambda_{\rm obs}$ (\AA) & Identification & Flux$_{\rm {X1}}$ & $\sigma_{\rm{X1}}$ & Flux$_{\rm{X2}}$ & $\sigma_{\rm{X2}}$ \\
\hline
14760   & 14761 & He~{\sc ii}? & 1.57 & 0.12  &  --  &  --      \\
14863   & 14863 & [Ti~{\sc i}]?  & 0.13 & 0.03 & -- & -- \\
14882   & 14880 & He~{\sc ii}?       & 0.33 & 0.04 & -- & -- \\
15083   & 15082 & H~{\sc i}\,4-22     & 0.11 & 0.02 & -- & -- \\
15133   & 15136 & H~{\sc i}\,4-21     & 0.14 & 0.03 & -- & -- \\
15192   & 15193 & H~{\sc i}\,4-20     & 0.19 & 0.03 & -- & -- \\
15261   & 15263 & H~{\sc i}\,4-19     & 0.14 & 0.02 & 0.03 & 0.02 \\
15335   & 15333 & [Fe~{\sc ii}]       & 0.12 & 0.03 & -- & -- \\
15342   & 15346 & H~{\sc i}\,4-18     & 0.19 & 0.03 & 0.08 & 0.01 \\
15439   & 15442 & H~{\sc i}\,4-17     & 0.32 & 0.03 & 0.14 & 0.01 \\
15557   & 15560 & H~{\sc i}\,4-16     & 0.42 & 0.05 & 0.12 & 0.024 \\
15701   & 15704 & H~{\sc i}\,4-15     & 0.46 & 0.04 & 0.10 & 0.01 \\
15716   & 15720  & [V~{\sc ii}]?            & 0.26 & 0.03 & 0.02 & 0.01  \\
15881   & 15884 & H~{\sc i}\,4-14     & 0.72 & 0.05 & 0.22 & 0.03 \\
16109   & 16112 & H~{\sc i}\,4-13     & 0.86 & 0.03 & 0.29 & 0.04 \\
16407   & 16410 & H~{\sc i}\,4-12     & 1.13 & 0.03 & 0.38 & 0.02 \\
16436   & 16436 & [Fe~{\sc ii}]       & --    & --  & 0.34 & 0.02 \\
16807   & 16809 & H~{\sc i}\,4-11     & 1.53 & 0.03 & 0.48 & 0.03 \\
16914   & 16919 & He~{\sc i}         & 0.28 & 0.04 & -- & -- \\
17002   & 17005 & He~{\sc i}         & 0.65 & 0.03 & 0.33 & 0.04 \\
17362   & 17365 & H~{\sc i}\,4-10    & 2.01 & 0.05 & 0.63 & 0.03 \\
18139   & 18142 & He~{\sc i}         & 0.37 & 0.05 & -- & -- \\
18174   & 18179 & H~{\sc i}\,4-9$^{\dagger}$ & 6.85 & 0.18 & 1.82 & 0.12 \\
18445   & 18454 & He~{\sc i}?  & 0.53  & 0.10 & --  & --  \\

\hdashline
19446  & 19443 & H~{\sc i} 4-8$^{\dagger}$ & 5.28 & 0.20 & 7.39 & 0.21 \\
20338  & 20334 & H$_2$\,1-0S(2)  & 0.63 & 0.05 & 0.67 & 0.19 \\
20581  & 20585 & He~{\sc i}      & 9.63 & 0.16 & 6.63 & 0.27 \\
21121  & 21125 & He~{\sc i}      & 0.98 & 0.10 & 0.70 & 0.07 \\
21218  & 21216 & H$_2$\,1-0S(1)  & 2.90 & 0.08 & 2.12 & 0.15 \\
21608  & 21613 & He~{\sc i}?     & 0.55 & 0.11 & 0.36 & 0.11 \\
21655  & 21658 & H~{\sc i}\,4-7 (Br$\gamma$) & 18.22 & 0.21 & 9.72 & 0.20 \\
22233  & 22233 & H$_2$\,1-0S(0)  & 0.77 & 0.12 & 0.56 & 0.05 \\
22477  & 22470 & H$_2$\,2-1S(1)  & 0.33 & 0.07 & 0.19 & 0.05 \\
22870  & 22866 & ?               & 0.66 & 0.07 & 0.14 & 0.06 \\
24030  & 24039 & He~{\sc i}?     & 0.48 & 0.16 & -- & -- \\
24066  & 24063 & H$_2$\,1-0Q(1)  & 2.56 & 0.25 & 0.73 & 0.15 \\
24134  & 24131 & H$_2$\,1-0Q(2)  & 0.46 & 0.10 & 0.52 & 0.13 \\
24237  & 24234 & H$_2$\,1-0Q(3)  & 2.12 & 0.20 & 1.09 & 0.15 \\
24548  & 24551 & H$_2$\,1-0Q(5)  & 0.80 & 0.15 & -- & -- \\

\hline
\end{tabular}
\tablefoot{No reddening correction has been applied to the line fluxes. Using the Cardelli+1989 extinction law with R$_{V}=$3.1 and adopting a extinction coefficient c$_{\beta}$, one obtains A$_{V}$ and then A$_{\lambda}$ for each line (see the Section~\ref{sect3} for details). \\
$^{\dagger}$ These emission lines are located at the edge of the spectral coverage; therefore, their measured fluxes are less reliable.
}
\end{table*}

\begin{table*}[ht]
\caption{Emission line fluxes ($\times 10^{-15}$ erg s$^{-1}$ cm$^{-2}$) in the LISs of NGC~7354. H-band lines are above the dashed line, K-band lines below. Line identifications are primarily based on the \href{https://linelist.pa.uky.edu/atomic/}{Atomic Line List} (air wavelengths), while H$_2$ transitions follow \citet{2011A&A...528A..74A} and \citet{1987ApJ...322..412B}.}
\centering
\small
\begin{tabular}{l l c c c c c c c c c}
\hline
$\lambda_{\rm rest}$ & $\lambda$ (\AA) & Identification & Flux$_{\text{J1}}$ & $\sigma_{\text{J1}}$ & Flux$_{\text{J2}}$ & $\sigma_{\text{J2}}$ & Flux$_{\text{J3}}$ & $\sigma_{\text{J3}}$ & Flux$_{\text{Jet}}$ & $\sigma_{\text{Jet}}$ \\
\hline
14753 & 14752 & [Fe~{\sc ii}]? & 0.90 & 0.28   & 0.60 & 0.33   & -- & -- & 1.25 & 0.45 \\  
15439 & 15438 & H~{\sc i}\,4-17 & 0.66 & 0.13  & 0.25 & 0.05   & --    & -- & 1.17 & 0.06 \\
15556 & 15556 & H~{\sc i}\,4-16 & 0.80 & 0.13  & 0.69 & 0.14   & --    & -- & 1.47 & 0.07 \\
15701 & 15702 & H~{\sc i}\,4-15 & 0.71 & 0.11  & 0.31 & 0.08  & --    & -- & 1.30 & 0.14 \\
15719 & 15718 & [V~{\sc ii}]?  & 0.15 & 0.07 & --    & --      & --    & -- & 0.29 & 0.09  \\
15855 & 15855 & He~{\sc i}?     & 0.20 & 0.06 & 0.21 & 0.08  & 0.11 & 0.03 & 0.35 & 0.05 \\
15881 & 15882 & H~{\sc i}\,4-14 & 1.02 & 0.10  & 0.39 & 0.09  & 0.06& 0.03 & 1.56 & 0.16 \\
16109 & 16111 & H~{\sc i}\,4-13 & 1.34 & 0.12  & 0.27 & 0.09  & --    & -- & 1.93 & 0.11 \\
16407 & 16409 & H~{\sc i}\,4-12 & 2.08 & 0.08 & 0.64 & 0.07  & 0.10 & 0.05 & 2.83 & 0.11 \\
16435 & 16446 & [Fe~{\sc ii}]?  & 0.22 & 0.05 & --    & --      & --    & --     & 0.36 & 0.06 \\
16807 & 16808 & H~{\sc i}\,4-11 & 2.18 & 0.15  & 0.70 & 0.09  & 0.16 & 0.03 & 3.56 & 0.11 \\
16899  & 16900 & [V~{\sc ii}]?       & 0.72 & 0.07  & --    & --      & -- & -- & 0.60 & 0.09 \\
16914 & 16920 & He~{\sc i} & 0.41 & 0.05 & --    & --      & -- & -- & 0.74 & 0.10 \\
17002 & 17005 & He~{\sc i} & 1.57 & 0.07 & 0.81 & 0.14   & 0.32 & 0.09 & 2.18 & 0.13 \\  
17362 & 17363 & H~{\sc i}\,4-10 & 3.97 & 0.11   & 1.35 & 0.07  & 0.29 & 0.06 & 6.58 & 0.19 \\
17880 & 17881 & H$_2$\,1-0S(6) & 1.05 & 0.10  & 0.43 & 0.10  & 0.39 & 0.08 & 1.26 & 0.26 \\
18139 & 18141 & He~{\sc i}  & 0.61 & 0.37   & --    & --      & --    & --    & 0.64 & 0.34 \\
18174 & 18177 & H~{\sc i}\,4-9$^{\dagger}$ & 15.48 & 0.70  & 4.76 & 0.30   & 1.43 & 0.24 & 24.26 & 0.67 \\
18243 & 18245 & [V~{\sc ii}]?$^{\dagger}$   & 1.62 & 0.38   & 0.62 & 0.17   & 0.89 & 0.20 & 2.69 & 0.37 \\
18358 & 18271 & H$_2$\,1-0S(5)?$^{\dagger}$ & 1.30 & 0.36 & 0.87 & 0.20   & 0.55 & 0.20 & 0.94 & 0.346 \\
18451 & 18454 & [V~{\sc iii}]$^{\dagger}$   & 1.07 & 0.30  & 0.52 & 0.29   & --    & --    & 2.87 & 0.43 \\

\hdashline
19414 & 19414 & He~{\sc i}?$^{\dagger}$ & 0.22 & 0.05 & --     & --     & --    & --     & 0.31 & 0.16 \\
19446 & 19444 & H~{\sc i} 4-8$^{\dagger}$ & 4.25 & 0.09 & 1.74  & 0.10 & 0.47 & 0.11  & 6.57 & 0.22 \\  
20275 & 20273 & He~{\sc ii}? & 0.31 & 0.04 & 0.25 & 0.04  & 0.20 & 0.06 & 0.69  & 0.12 \\
20338 & 20336 & H$_2$\,1-0S(2) & 0.77 & 0.07 & 0.66 & 0.06  & 0.33 & 0.07  & 1.67  & 0.14 \\
20581 & 20585 & He~{\sc i}     & 3.87 & 0.10 & 1.62  & 0.05  & 0.28 & 0.06  & 5.23  & 0.19 \\
20735 & 20733 & H$_2$\,2-1S(3) & 0.53 & 0.06 & 0.34 & 0.05  & 0.23 & 0.06  & 0.87  & 0.13 \\
21121 & 21125 & He~{\sc i} & 0.36 & 0.04 & 0.15 & 0.04  & --   & --    & 0.52 & 0.08  \\
21132 & 21140 & He~{\sc i}? & 0.26 & 0.04 & 0.09 & 0.04  & --   & --   & 0.38 & 0.08  \\
21218 & 21217 & H$_2$\,1-0S(1) & 2.26 & 0.05 & 1.65  & 0.05 & 0.85  & 0.05 & 4.50 & 0.11 \\
21608 & 21613 & He~{\sc i}?   & 0.40 & 0.06 & 0.10 & 0.03  & --    & --      & 0.58 & 0.06 \\
21661 & 21658 & H~{\sc i}\,4-7 (Br$\gamma$) & 14.65 & 0.11 & 4.67 & 0.05 & 1.48 & 0.06 & 20.63 & 0.18 \\
21885 & 21886 & He~{\sc ii}? & 1.22 & 0.04 & -- & -- & -- & -- & 1.16 & 0.12 \\
22233 & 22229 & H$_2$\,1-0S(0) & 0.46 & 0.05 & 0.48 & 0.04 & 0.27 & 0.06 & 0.98 & 0.09 \\
22477 & 22474 & H$_2$\,2-1S(1) & 0.26 & 0.03 & 0.25 & 0.03 & 0.15 & 0.04 & 0.63 & 0.11 \\
22870 & 22863 & ? & 1.06 & 0.03 & 0.34 & 0.04 & 0.16 & 0.06 & 1.46 & 0.07 \\
24030 & 24031 & He~{\sc i}  & 0.24 & 0.07 & 0.22 & 0.07 & --    & --    & 0.42 & 0.17  \\
24066 & 24064 & H$_2$\,1-0Q(1) & 1.15 & 0.10 & 1.77 & 0.10 & 0.90 & 0.08 & 3.48 & 0.22 \\
24134 & 24131 & H$_2$\,1-0Q(2) & 0.43 & 0.08 & 0.70 & 0.08 & 0.29 & 0.06 & 1.14 & 0.17 \\
24237 & 24232 & H$_2$\,1-0Q(3) & 1.17 & 0.10 & 1.76 & 0.10 & 1.12 & 0.09 & 3.91 & 0.23 \\
24307 & 24311 & H~{\sc i}\,5-20 & 0.37 & 0.08 & 0.14 & 0.08 & --   & --     & 0.67 & 0.17 \\
24375 & 24371 & H$_2$\,1-0Q(4) & 0.25 & 0.07 & 0.47 & 0.08 & 0.37 & 0.07 & 1.04 & 0.17 \\
24483 & 24481 & H~{\sc i}\,5-19 & 0.34 & 0.08 & --    & --     & --    & --    & 0.71 & 0.16  \\
24548 & 24546 & H$_2$\,1-0Q(5) & 0.89 & 0.09 & 1.42  & 0.09  & 0.81 & 0.08 & 2.89 & 0.20 \\
24693 & 24696 & H~{\sc i}\,5-18 & 0.97 & 0.20 & --     & --      & --    & --     & 1.27 & 0.33 \\
\hline
\end{tabular}
\tablefoot{No reddening correction has been applied to the line fluxes. Using the Cardelli+1989 extinction law with R$_{V}=$3.1 and adopting a extinction coefficient c$_{\beta}$, one obtains A$_{V}$ and then A$_{\lambda}$ for each line (see the Section~\ref{sect3} for details). \\
$^{\dagger}$ These emission lines are located at the edge of the spectral coverage; therefore, their measured fluxes are less reliable.
}

\end{table*}

\twocolumn

\begin{table}[ht]
\centering
\caption{Ionized masses (M$_{\rm ion}$) derived from Br$\gamma$ emission lines, warm molecular masses (M$_{\rm warm}$) derived from H$_2$~1--0\,S(1) emission lines, and warm molecular masses estimated from H$_2$ column densities (M$_{\rm warm~column}$) obtained through the excitation diagrams (Sect.~\ref{sect:h2diag}), for all the LISs analyzed in this work. All masses are expressed in solar mass units.}
\label{tab:masses}
\small
\begin{tabular}{lccc}
\hline
LIS  &   M$_{\rm ion}$            &  M$_{\rm warm}$                  &  M$_{\rm warm~column}$  \\
     &   [$\times 10^{-04}$ M$_{\odot}$]  & [$\times 10^{-07}$ M$_{\odot}$]  & [$\times 10^{-07}$ M$_{\odot}$] \\
\hline
X1   &   19.07 $\pm$  1.38 &  13.15 $\pm$ 0.36 &  5.94 $\pm$ 0.34  \\
X2   &   10.18 $\pm$  0.76 &   9.61 $\pm$ 0.68 &  3.69 $\pm$ 0.95  \\
J1   &   22.55 $\pm$  7.36 &   7.64 $\pm$ 0.16 &  2.68 $\pm$ 0.34  \\
J2   &    7.19 $\pm$  2.35 &   5.58 $\pm$ 0.18 &  1.96 $\pm$ 0.29  \\
J3   &    2.27 $\pm$  0.75 &   2.88 $\pm$ 0.18 &  0.94 $\pm$ 0.07  \\
Jet  &   22.49 $\pm$  6.32 &  15.20 $\pm$ 0.37 &  3.16 $\pm$ 0.40  \\
\hline
\end{tabular}
\end{table}

\begin{figure}[ht]
       \centering
       \includegraphics[width=0.98\columnwidth]{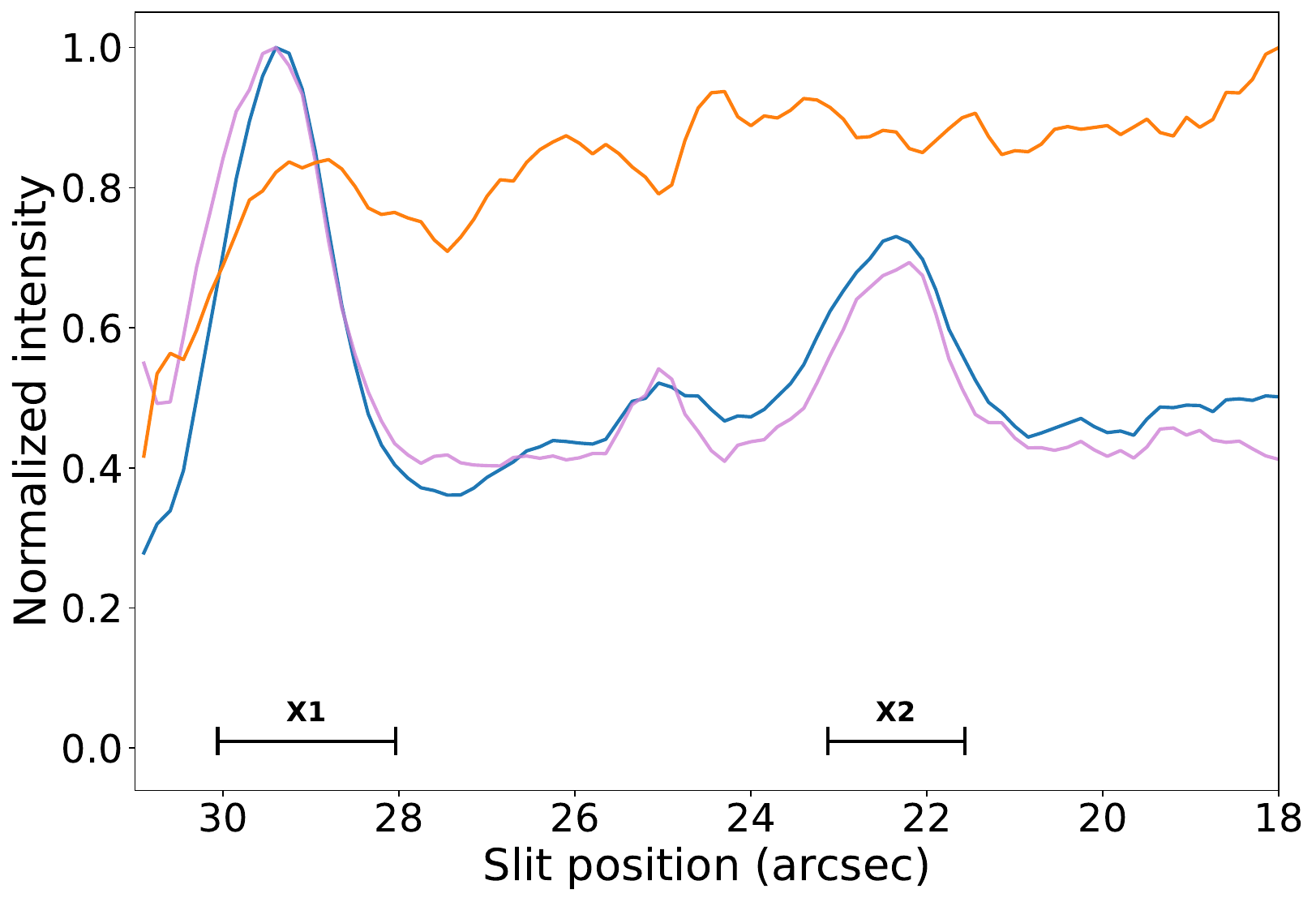}
       \includegraphics[width=0.98\columnwidth]{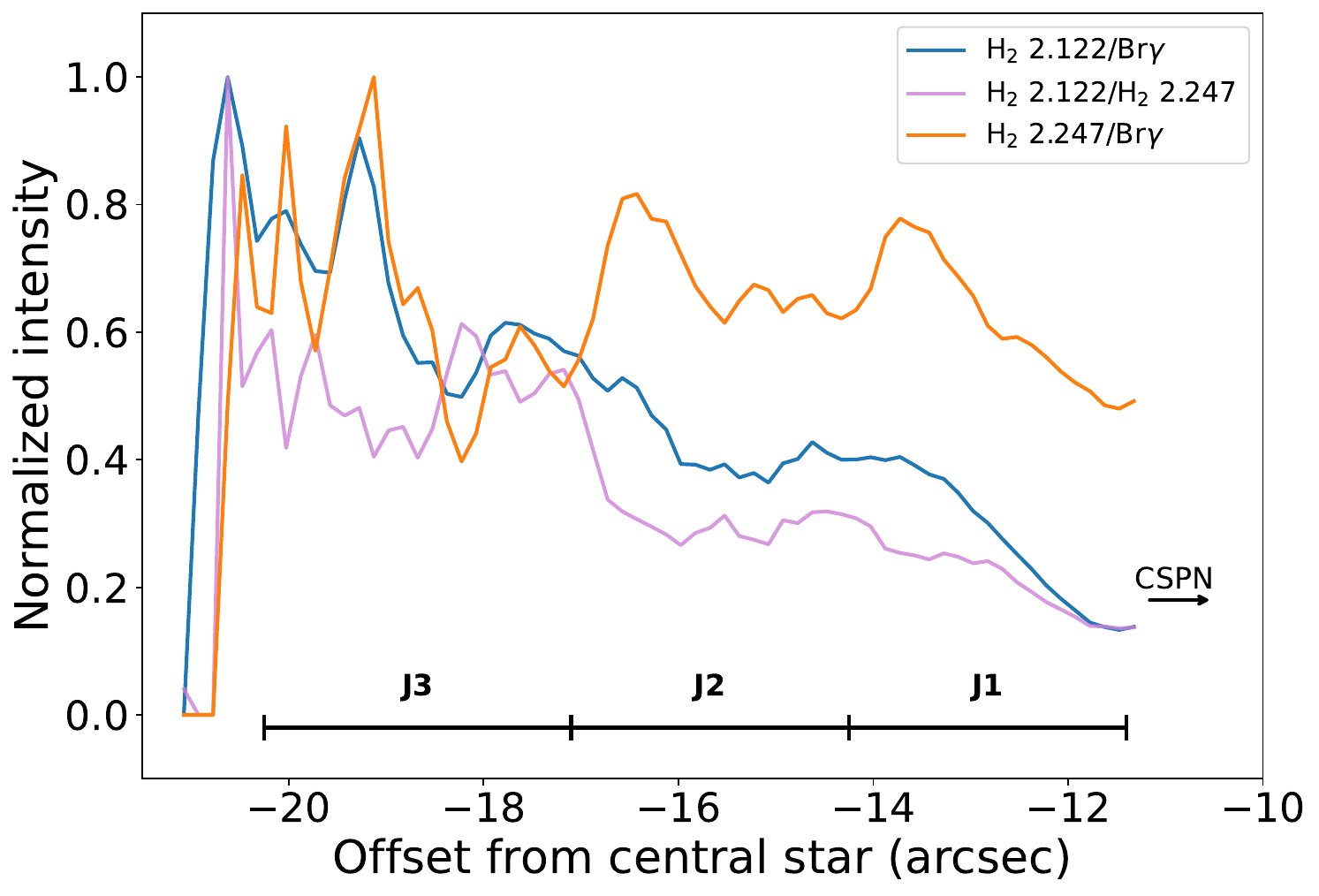}
       \caption{
       Spatial variation of the H$_2$ 2.122/Br$\gamma$, H$_2$ 2.247/Br$\gamma$ and H$_2$ 2.122/2.247 line ratios along the slit for NGC~6818 (\textit{top}) and NGC~7354 (\textit{bottom}). All profiles are normalized to their respective peak value.        }
       \label{fig:spatial_profiles2}%
\end{figure}

\section{Estimate of the total gas density}\label{appendixB}

As a complementary, independent check on the gas densities discussed in Sect.~\ref{sect:h2diag}, 
we estimate the total hydrogen density (atomic + molecular) using the empirical relation between 
total gas column density and interstellar reddening established by \citet{1978ApJ...224..132B}:

\begin{equation}
\hspace{1cm} N_{\rm total}({\rm H\,I}+{\rm H}_2) = 5.8\times10^{21}\,E(B-V)\ \ [{\rm cm}^{-2}].
\label{eq:bohlin}
\end{equation}

The extinction coefficients $c_\beta$ (Sect.~\ref{sect3}) were converted to $E(B-V)$ via 
$E(B-V)=c_\beta/1.47$, consistent with $R_V=3.1$ \citep[][]{1989ApJ...345..245C}. Assuming that hydrogen in these structures is predominantly molecular, $N({\rm H}_2)=0.5\,N_{\rm total}$  (warm+cold), and dividing by the path length $L$ of each structure along the line of sight --~adopted here as the projected length of each extraction along the slit (Table~\ref{tab:density_bohlin})~-- we obtain $n({\rm H}_2)=N({\rm H}_2)/L$ and, correspondingly, $n_{\rm total}=2\,n({\rm H}_2)$.

This estimate relies on two assumptions that should be kept in mind: (i) that the gas-to-dust ratio 
in these structures is comparable to that of the average diffuse ISM, from which 
Eq.~\ref{eq:bohlin} was derived, and (ii) that hydrogen is essentially fully molecular. Neither 
assumption is directly verified for gas in LISs, so the resulting 
densities (Table~\ref{tab:density_bohlin}) should be regarded as order-of-magnitude estimates.

\begin{table}[ht]
\centering
\caption{Total hydrogen density estimated from the gas-to-extinction relation of 
\citet{1978ApJ...224..132B} (Eq.~\ref{eq:bohlin}).}
\label{tab:density_bohlin}
\small
\begin{tabular}{lccc}
\hline
LIS & E(B-V)  & L          & $n_{\rm total}$  \\
    &         & (arcsec)   & [cm$^{-3}$] \\
\hline
X1    & 0.252 & 1.56 & 2.18$\times10^{4}$ \\
X2    & 0.252 & 2.02 & 1.69$\times10^{4}$ \\
J1    & 1.177 & 2.85 & 7.62$\times10^{4}$ \\
J2    & 1.177 & 2.85 & 7.62$\times10^{4}$ \\
J3    & 1.177 & 3.15 & 6.90$\times10^{4}$ \\
Jet   & 1.177 & 8.89 & 2.44$\times10^{4}$ \\
\hline
\end{tabular}
\end{table}

It is worth noting that the adopted path length L --~the projected extent of each extraction along the slit~-- represents a plausible but not unique choice. For the jet, in particular, this length is measured \textit{along} its elongated axis and does not constrain its (unresolved) transverse depth. If, instead, the slit width (0.45\arcsec, fully containing the jet's transverse extent; Sect.~\ref{obs}) is adopted as a proxy for this depth, the inferred density increases by approximately one order of magnitude, reaching $n_{\rm total}\sim4.83\times10^{5}$~cm$^{-3}$ and identical for all three jet segments, since in that case the geometrical factor no longer varies along the jet. 
The two approaches should therefore be regarded as bracketing the likely range of 
$n_{\rm total}$ for the jet ($\sim2$--$8\times10^{4}$~cm$^{-3}$ using the along-slit length, up to $\sim5\times10^{5}$~cm$^{-3}$ using the slit width), reflecting the uncertainty introduced by the lack of spatially resolved transverse information. 
The knots in NGC~6818, in contrast, are crossed by the slit roughly perpendicular to their filamentary major axis --~opposite to the jet's configuration~-- so that the along-slit extraction samples their transverse width rather than their full length, yielding a comparatively more representative estimate of the line-of-sight depth and, correspondingly, a smaller ambiguity in $n_{\rm total}$.

\end{appendix}

\end{document}